\documentstyle[12pt,epsf,epsfig]{article}

\newcommand{\bcen}{\begin{center}}
\newcommand{\ecen}{\end{center}}

\newcommand{\beq}{\begin{equation}}
\newcommand{\eeq}{\end{equation}}

\newcommand{\beqn}{\begin{eqnarray}}
\newcommand{\eeqn}{\end{eqnarray}}

\newcommand{\bdm}{\begin{displaymath}}
\newcommand{\edm}{\end{displaymath}}

\begin{document}
\begin{flushright}
{\large UGI-00-14}
\end{flushright}
\begin{center}
{\Large \bf The $\rho$ Spectral Function in a Relativistic Resonance Model}
\footnote{Work supported by BMBF and DFG.}
\bigskip
\bigskip

{M. Post, S. Leupold and U. Mosel}

\bigskip
\bigskip
{ \it
Institut f\"ur Theoretische Physik, Universit\"at Giessen,\\
D-35392 Giessen, Germany
}
\end{center}

\bigskip
\bigskip


\begin{abstract}

We calculate the spectral function $A_\rho$ of the $\rho$ meson in nuclear matter.
The calculation is performed in the {\it low density} approximation, where the 
in-medium self energy $\Sigma_{med}$
is completely determined by the vacuum $\rho\,N$ forward
scattering amplitude. This amplitude is derived from a relativistic resonance model.
In comparison to previous non-relativistic calculations we find a much weaker
momentum dependence of $\Sigma_{med}$, especially in the transverse channel.
Special attention is paid to uncertainties in the model. Thus, we compare 
the impact of different 
coupling schemes for the $RN\rho$ interaction on the results and discuss 
various resonance parameter sets.\\


 
\end{abstract}


\newpage

\section{Introduction}

The question of how vector meson properties change at finite baryon 
density or temperature has triggered a lot of theoretical and experimental work 
in the last decade. 
Of particular interest are the in-medium properties of $\rho$ and $a_1$ meson, 
which are chiral partners. 
One of the most evident signatures of the spontaneous breaking of 
chiral symmetry in vacuum is the large mass difference between
these mesons. 
The claim is that the in-medium mass spectra of $\rho$ and $a_1$ meson can be related to 
the partial restauration of chiral symmetry.
If a chirally symmetric phase is reached 
at high density/temperature, a degeneracy of their spectral functions
is expected \cite{vk,ka}. 
Whereas only a few studies exist on the in-medium properties of the $a_1$ meson 
\cite{pi,ch},
numerous theoretical works have been performed for the $\rho$ meson
\cite{brownrho,hl,leu,fri,fp,pp98,fri2,weise,chanfray,herrmann,urban,rw,rw2}.

It has been argued by several groups that 
the mass of the $\rho$ meson is related to the 
expectation value of the scalar quark condensate. The latter is supposed to 
decrease at finite nuclear density due to the restauration of chiral symmetry, 
thereby inducing a mass shift on the $\rho$ meson.
The most famous of these models is that of Brown and Rho \cite{brownrho}. It predicts a 
downward shift for the $\rho$ meson mass of about $15\% - 20\%$ at normal nuclear 
matter density. Similar results have been found in \cite{hl} using the QCD sum rule
approach under the assumption that the width of the $\rho$ meson remains small in
the nuclear medium (cf. \cite{leu} for a thorough discussion of that point).

In other approaches the properties of the $\rho$ meson in nuclear
matter are calculated by taking into account conventional nuclear many body effects. 
At low nuclear densities $\rho_N$
the self energy $\Sigma_{med}$ can be calculated in the $\rho_N\,T$
approximation ({\it low density} approximation) and is  
completely determined by the knowledge of the vacuum $\rho\,N$ forward scattering 
amplitude $T$. Since this quantity is not directly accessible by experiment, one
needs to construct a model which incorporates experimental constraints
as far as possible. This was first pointed out in \cite{fri}.
Information on $\rho N$ scattering comes mostly from
$\pi N \rightarrow \pi \pi N$ processes. One of the most successful descriptions
of these reactions has been achieved by Manley {\it et al}
within an isobar model \cite{man1,man2}.

A resonance model for the $\rho\,N$ scattering amplitude allows the incorporation of 
experimental constraints in a straightforward fashion. 
Such a strategy was first suggested in \cite{fp}, where 
only $p-$wave resonances were taken into account. However, as shown in 
our previous paper \cite{pp98}, the $D_{13}(1520)$ 
resonance plays a dominant role in 
$\rho N$ scattering and can therefore not be neglected in a complete analysis.
In both models the resonance parameters were taken from the PDG \cite{pdg} and the 
calculation was performed within a non-relativistic framework.
An alternative approach has been chosen in \cite{fri2}, where the baryonic resonances
are not introduced as excited three-quark states, but
are generated dynamically in a coupled channel approach from meson-nucleon 
and meson-$\Delta$ rescattering.
In \cite{weise} the $\rho N$ scattering amplitude is derived from an effective 
chiral Lagrangian theory, neglecting, however, resonant contributions.

Going beyond the {\it low density} approximation, various groups have considered 
the change in the 2 $\pi$ decay of the $\rho$ as induced by the modified
$\pi$ spectrum in the nuclear medium, see e.g. 
\cite{chanfray,herrmann,urban,rw,rw2}. In \cite{rw,rw2} resonance 
contributions have also been included. 

Qualitatively, all these models predict a strong broadening of 
the $\rho$ meson. The mass, however, remains nearly unchanged. This finding is in
good agreement with QCD sum rules, see e.g. \cite{leu,weise}.

So far experimental signals indicating a change
of the $\rho$ spectrum in the nuclear medium, 
have been obtained by the CERES 
\cite{ceres1,ceres2,ceres4} and the HELIOS collaboration \cite{ceres3}. 
In these experiments the dilepton spectra display a downward-shift of strength  
in the region of the $\rho$ meson
as compared to theoretical predictions.
Agreement between theory and experiment can be reached
by introducing medium modifications for the $\rho$ meson \cite{ca}.
On the other hand, in \cite{koch} the claim is made, that the CERES data do not
necessarily hint to a medium modification of the $\rho$ meson, but can also be 
explained with its vacuum properties. 

Certainly, in order to gain a more quantitative understanding of the $\rho$   
meson properties in nuclear matter,
further theoretical studies, especially on the momentum dependence of 
the spectral function, and improved experimental statistics,
are mandatory. 
The sensitivity on details of the spectral function is
even enhanced in photo-nuclear reactions, as was demonstrated in \cite{effe}.

Up to now all existing calculations \cite{fp,pp98,rw,rw2}
have treated the resonance contribution
in a non-relativistic manner. As we will
show, this introduces substantial uncertainties concerning the momentum 
dependence of the $\rho$ spectral function.
We therefore do not only lift the non-relativistic approximation, 
but also study the dependence of the results on various coupling schemes.
Furthermore, we analyse the influence of different resonance parameter sets
on the results. We thus examine up to which
extent safe conclusions about the propagation of $\rho$ mesons in nuclear matter
can be drawn from a resonance model.

The paper is organized as follows.
In Sect. \ref{overview} we will present an overview of the formalism employed
to describe the spectral function. Furthermore, an arbitrariness 
in the non-relativistic approach is exposed 
and thus the need for a relativistic
model is motivated. 
The basic ingredients of the relativistic
calculation will be introduced in Sect. \ref{rel}.
A problem related with the description of the resonance propagator and its solution 
is presented in Sect. \ref{resprop}. 
In Sect. \ref{manley}
we analyze the quality and properties of different resonance parameter sets and their
influence on the results. 
In Sect. \ref{results} the results of the relativistic calculation are presented
and then compared to the non-relativistic calculation. The 
impact of various coupling schemes for the vertex $RN\rho$ is studied in Sect. 
\ref{uncertain}. We summarize our results in Sect. \ref{conclu}.


\section{Overview of the Model}
\label{overview}

In this section we 
explain the formalism on which our calculation of the in-medium spectral function
of the $\rho$ meson is based and introduce some basic notation.
We then review the non-relativistic scheme which has been employed in all
works up to now \cite{fp,pp98,rw,rw2} and use some simple kinematical considerations 
to point out why a fully relativistic framework is needed.

\subsection{Definition of the Basic Quantities}
\label{basics}

The spectral function is defined as the imaginary part of the 
propagator. Due to Lorentz invariance the vacuum spectral function 
of the $\rho$ meson depends only on one kinematic variable --
its invariant mass --  and both transverse and longitudinal modes have the 
same spectral distribution.
The presence of nuclear matter breaks Lorentz invariance. Consequently
transverse and longitudinal propagation modes of the $\rho$ meson
receive different in-medium modification. Also, the corresponding
spectral functions depend both on
the energy $\omega$ of the $\rho$ meson and its three-momentum ${\bf q}$:
\beq
A_\rho^{T/L}(\omega,{\bf q}) = \frac{1}{\pi}\frac{\mbox{Im } \Sigma^{T/L}(\omega,{\bf q})}
{(\omega^2-{\bf q}^2 - m_\rho^2+\mbox{Re } \Sigma^{T/L}(\omega,{\bf q}))^2+
\mbox{Im } \Sigma^{T/L}(\omega,{\bf q})^2}   \qquad . 
\eeq
$m_\rho$ is the pole mass of the $\rho$ meson. $\Sigma^{T/L}$ is the 
transverse or longitudinal part of the self energy 
of the $\rho$ meson and can be decomposed into a vacuum and an in-medium part:
\beq
\Sigma^{T/L}(\omega,{\bf q}) = \Sigma_{vac}(q^2) + 
\Sigma_{med}^{T/L}(\omega,{\bf q}) \qquad .
\eeq

The imaginary part of the former is to lowest order in the coupling constant 
given by the 2-pion decay width of the $\rho$ meson \cite{weise,herrmann}:
\beq
{\mbox{Im }}\Sigma_{vac}(q^2) = \sqrt q^2 \,\Gamma_{\pi\pi}(q^2) \qquad,
\eeq
and
\beq
\Gamma_{\pi\pi}(q^2) = \frac{m_\rho^2}{q^2}\,\Gamma_0\,
\left(\frac{{\bf p}}{{\bf p}_{m_\rho}} \right)^3 \qquad .
\eeq
Here $q^2$ is the squared invariant mass of the $\rho$ meson, $\Gamma_0$ its decay width
on the pole mass and ${\bf p}_{m_\rho}\,({\bf p})$ 
denotes the 3-momentum of the pions measured
in the rest frame of a decaying $\rho$ meson with mass $m_\rho\,(\sqrt q^2)$. 
For invariant masses of the $\rho$ meson
up to around $1$ GeV - which we consider in this paper - the real part of 
$\Sigma_{vac}(q^2)$ shows only small variations and can be neglected \cite{weise}. 

At finite nucleon density $\rho_N$ the in-medium self energy $\Sigma_{med}^{T/L}$ 
is given by the following expression:
\beq
\label{sigma}
\Sigma_{med}^{T/L}(\omega,{\bf q}) = \int_{\Omega}\frac{d^3{\bf p_N}}{(2\pi)^3\,2 E_N}\,
T_{tot}^{T/L}(q,p_N)
\eeq
In our notation $q=(\omega,{\bf q})$ 
and $p_N=(E_N,{\bf p}_N)$ are the 4-vectors of $\rho$ meson and nucleon,
respectively. The integration is restricted to the volume of the Fermi volume,
denoted here by $\Omega$. By performing the integration in the rest frame of nuclear matter
$\Omega$ is simply the Fermi sphere, 
and the ${\bf {p}_N}$ integration reaches from $0$ to the 
Fermi momentum ${\bf p_F}$.
$T_{tot}^{T/L}$ is the $\rho N$ forward scattering 
amplitude, as discussed below. %
Note that $\Sigma_{med}(\omega,{\bf q})$ is a Lorentz invariant quantity, since
both the forward scattering amplitude $T_{tot}^{T/L}$ and the integrations measure
$\frac{d^3{\bf p_N}}{(2\pi)^3\,2 E_N}$ are Lorentz scalars.

At sufficiently low nuclear densities it is advantageous to expand 
the self energy in terms of the Fermi momentum. This is most easily done
in the rest frame of nuclear matter. The 4-momentum of the nucleon
simply becomes $p_N=(m_N,0)$.
One then recovers the $\rho_N \,T$ or {\it low density} approximation
\cite{dover}:
\beq
\label{lowdens}
\Sigma_{med}^{T/L}(q) = \frac{1}{8 m_N}\,\rho_N\,T_{tot}^{T/L}(q,p_N) \quad .
\eeq

\begin{figure} 
\begin{center}
\epsfig{file=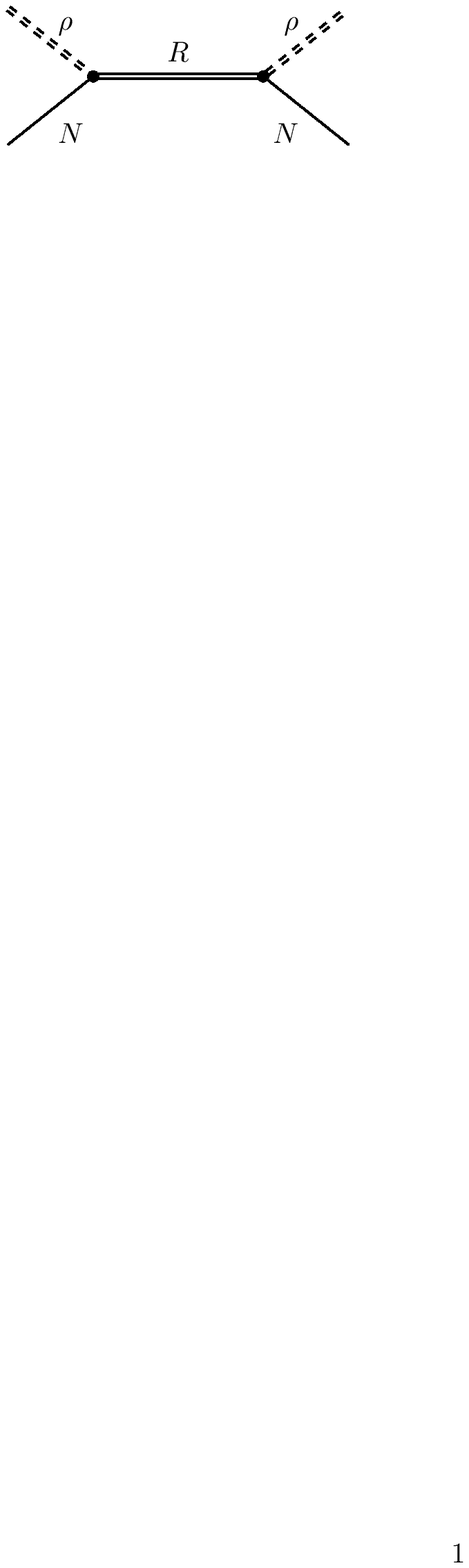,width=7cm}
 \caption{ 
\label{rho_n} Resonance contribution to the $\rho\,N$ scattering amplitude.}
\end{center}
\end{figure}

The forward scattering amplitude $T_{tot}^{T/L}$ is obtained as a sum over all
included resonances. For a single resonance the scattering amplitude is depicted in 
Fig. \ref{rho_n}. The corresponding contribution to the transverse and longitudinal
channel $T^{T/L}$ is obtained from the 
tensor $T^{\mu\nu}(q,p_N)$ upon contracting with the projection operators
$P^{T/L}_{\mu\nu}$ \cite{pp98,weise,rw}:
\beqn
\label{selfen1}
T^T &=& \frac{1}{2}\,T^{\mu\nu}\,P^T_{\mu\nu} \nonumber \\ && \\
T^L &=& T^{\mu\nu}\,P^L_{\mu\nu} \qquad .\nonumber 
\eeqn
The general form of $T^{\mu\nu}(q,p_N)$ reads:
\beq
\label{selfen0}
T^{\mu\nu}(q,p_N) = \left\{
\begin{array}{ll}
 f^2\,Tr \, {\cal V}_\frac{1}{2}^\mu(q,p_N) \, D_{\frac{1}{2}}(k)\,
{\cal V}_\frac{1}{2}^\nu(q,p_N) \,(\slash \hspace{-0.2cm}p + m_N)
 &\mbox{for}\,\, J=\frac{1}{2} \, , \\ \\ 
 f^2\,Tr \, {\cal V}_\frac{3}{2}^{\mu\alpha}(q,p_N) 
\, D_{\frac{3}{2}\,\,\alpha\beta}(k)\,
{\cal V}_\frac{3}{2}^{\beta\nu}(q,p_N) \, (\slash\hspace{-0.2cm}p + m_N)
&\mbox{for}\,\, J=\frac{3}{2} \, ,
\end{array} \right.
\eeq
where $k=p_N+q$ is the four-momentum of the intermediate resonance.
The coupling strength at the $RN\rho$ vertex is given by $f$.
A detailed discussion of both the relativistic vertex functions
${\cal V}_\frac{1}{2}^\mu(q,p_N)$ and ${\cal V}_\frac{3}{2}^{\mu\nu}(q,p_N)$ 
as well as the  relativistic propagators of spin-$\frac{1}{2}$ and 
spin-$\frac{3}{2}$ resonances 
$ D_{\frac{1}{2}}(k)$ and $ D_{\frac{3}{2}}^{\alpha\beta}(k)$ will 
follow in Sect. \ref{rel}.

\begin{figure} 
\begin{center} \epsfig{file=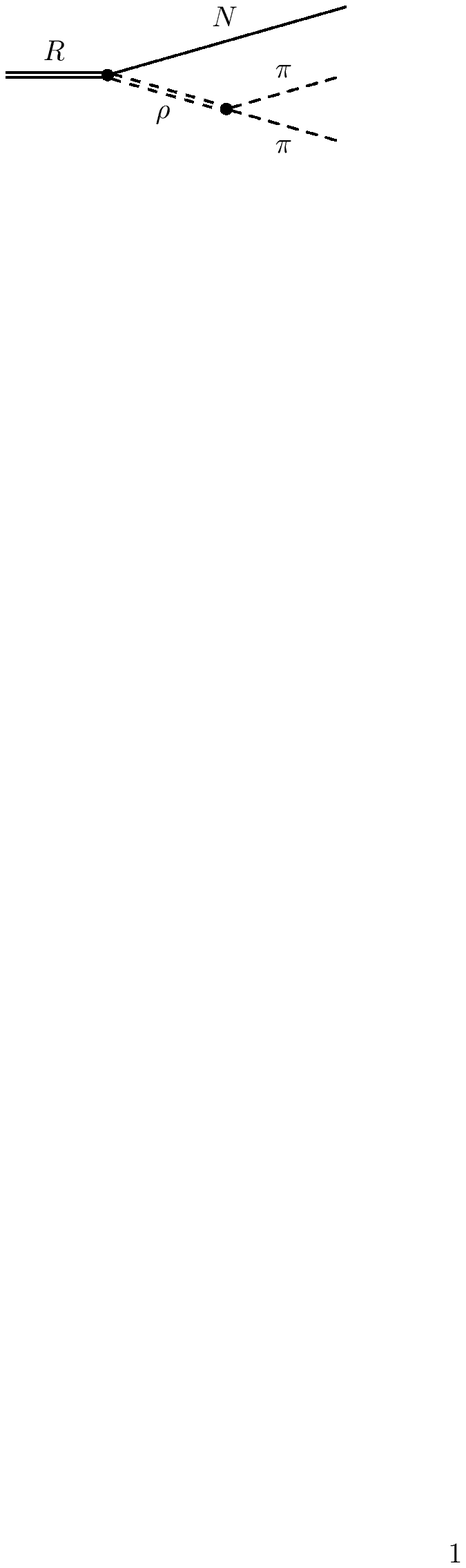,width=7cm} 
 \caption{ 
\label{decay} Decay of a resonance into two pions via an intermediate $\rho$ meson.}
\end{center}
\end{figure}

The coupling constant $f$ is obtained from a fit 
to the measured partial decay width $R\rightarrow N\,\rho$. 
In Sect. \ref{manley} we discuss in detail from which analysis the
resonance parameters are taken. Since many of the
resonances which are considered in this work are below or relatively
close to the nominal threshold for this decay of 
$\sqrt s = m_N + m_\rho \approx 1.7$ GeV,
we integrate over the vacuum spectral function of the $\rho$ meson 
(see Fig. \ref{decay}). The decay width then reads: 
\beqn
\label{coupldet}
\Gamma_{N\rho}(s) &=& \frac{1}{2\,j_R+1}\,\frac{I_\Gamma}{8\,\pi\,s}\,
\int\limits_{2m_\pi}^{\sqrt{s}-m_N}\!\!\!\!dm \,2m\,F(s)^2 \, \nonumber \\ &&  \qquad \quad
\times \,A_\rho^{vac}(m^2)\,(2\,|{\cal M}_{RN\rho}^T|^2
+|{\cal M}_{RN\rho}^L|^2) \, {\bf q_{cm}}\qquad .
\eeqn
The spin of the resonance is denoted by $j_R$, $I_\Gamma$
is an isospin factor. $A_\rho^{vac}$ is the vacuum spectral function of the 
$\rho$ meson. By ${\bf q_{cm}}$ we denote the 3-momentum of 
the $\rho$ meson in the rest frame of the decaying resonance. 
$F(s)$ is a formfactor that will be explained in Eq. \ref{formfac}. 
The squared matrix element for the resonance decay $|{\cal M}_{RN\rho}^{T(L)}|^2$
reads in terms of the vertex functions $\cal V$ (here for spin-$\frac12$ resonances):
\beq 
\label{matrix2}
|{\cal M}_{RN\rho}^{T(L)}|^2 = -f_{RN\rho}^2\, 
\frac{P^T_{\mu\nu}}{2}\,\left( P^L_{\mu\nu}\right)\,
\sum_{s,r} {\bar u^R}_r(k)\,{\cal V}^{\mu}_{\frac12}
\,u_s(p_N)\,{\bar u_s}(p_N)\,{\cal V}^{\nu}_{\frac12}\,u^R_r(k) \qquad.
\eeq
The spinors $u_r^R(k)$ and $u_s(p_N)$ denote resonance and nucleon spinors respectively.
The numerical values for the coupling strength $f_{RN\rho}$, 
which we obtain this way, are given 
in Table \ref{barres}. 
We have defined $\Gamma_{N\rho}$ for an arbitrary invariant mass  
$\sqrt s$ of the resonance, since 
we use Eq. \ref{coupldet} also as a parameterization of the mass dependence of
the decay width.

The $\rho_N\,T$ approximation Eq. \ref{lowdens}
only describes two-body processes in the
nuclear medium. At sufficiently large densities also higher
order effects need to be taken into account -- currently, however,  a scale that
pins down the breakdown of the $\rho_N\,T$ approximation is not at hand.
A comprehensive description of many-body effects is very hard to obtain
\cite{pp98,rw2}.
Therefore, we incorporate only two typical in-medium corrections to the
low-density approximation:
\begin{itemize}
  \item integration over the Fermi sphere of the nucleons,
  \item renormalization of $\Sigma_{med}$ due to short range correlations.
\end{itemize}
In the case of $p$-wave resonances,
the short range correlations are incorporated via the Landau-Migdal parameters:
\beq
\Sigma_{med}(\omega,{\bf q}) = \frac{\Sigma_{med}}{1-g^\prime
\frac{\Sigma_{med}}{{\bf q}^2}}
\qquad .
\eeq
Only the $P_{33}(1232)$ is renormalized, leading to a suppression
of its contribution. For the Landau-Migdal parameter we take $g^\prime = 0.5$. For higher
$p$-wave resonances this parameter is not known. The actual calculation shows
very little effect of $g^\prime$ on the results. 
For $s$-wave resonances there is no information available on how to include 
short range correlations; 
thus we do not renormalize the contribution from these resonances.

\subsection{Critical Review of the Non-Relativistic Model}
\label{review}

We quickly review here the qualitative features 
of the results of a non-relativistic approach 
as found in our previous work \cite{pp98}. 
At low momenta the results are completely
determined by the properties of the excited resonances. 
Their quantum numbers 
and experimentally observed coupling strengths to the $\rho\,N$ channel provide
the necessary information.
Close to threshold the resonance contribution scales like ${\bf q}^{2\,l}$,
where ${\bf q}$ is the 3-momentum of the $\rho$ meson. 
Consequently, at small values of ${\bf q}$ $s$-wave resonances ($P=-1$)
yield finite contributions, whereas the contribution from  $p$-wave 
resonances ($P=+1$) is negligible.
Quantitatively, the results of our explicit calculation of 
$\Sigma_{med}$ \cite{pp98} reveal that the propagation of a $\rho$ meson in 
nuclear matter is dominated by the $s$-wave resonances $D_{13}(1520)$ 
and $D_{33}(1700)$ as a direct consequence of their large coupling to the 
$\rho\,N$ channel as given in \cite{pdg}. The spectral function 
receives a rich structure:
a new peak arises at an invariant mass of about $0.5$ GeV from the excitation 
of the $D_{13}(1520)$ and the peak at the pole mass of the $\rho$ meson
is broadened by the $D_{33}(1700)$.

Whereas at low momenta the main uncertainty of the results comes from 
the values of the coupling constants, at high momenta an additional model
dependence is introduced by the momentum dependence of the self energy.
In a non-relativistic calculation $p$-wave resonances do not couple to  
longitudinal $\rho$ mesons. Furthermore, $A^T$ and $A^L$ display a
completely different behaviour at high momenta. For $A^L$ the effects from 
resonance excitation are rather small and the spectral function is strongly
peaked at the $\rho$ mass.
$A^T$, in contrast, is completely smeared out in this kinematic regime. 

We will demonstrate now that the momentum dependence of the self energy
can not be reliably determined within a non-relativistic calculation
and we argue that therefore a relativistic framework is imperative in order to
put the results on a more solid basis.

The forward scattering amplitude is proportional to $f^2 \,{\cal V}^2$, see Eq.
\ref{selfen0}. From Eqs. \ref{coupldet} and \ref{matrix2} it is clear that the 
vertex function ${\cal V}$ enters also into the determination
of the the coupling constant $f$. It thus appears twice in the forward 
scattering amplitude.
An ambiguity now arises in the choice of the reference frame in which 
the non-relativistic matrix element is evaluated. 
In all previous works
\cite{fp,pp98,rw,rw2} the following choice was made: the coupling 
constant was obtained by evaluating the matrix element
in the cm-frame (rest frame of the resonance) 
whereas the self energy was calculated in the lab-frame (rest frame of nuclear matter).
This is a suggestive choice since the decay width is usually 
defined in the rest frame of the decaying particle whereas
a nuclear matter calculation is preferably performed in the rest frame 
of nuclear matter.
The non-relativistic reduction itself does not uniquely determine the
reference frame, as it assumes that both the resonance and nucleon
momentum are small compared to their masses. This condition is equally
well met in the cm-frame and in the lab-frame.

As shown in \cite{fp,pp98}, for a resonance of positive parity both the  
resonance decay width and the forward scattering amplitude 
are proportional to ${\bf q}^2$. For a given 
invariant energy $\sqrt s$ the three-momenta in lab-frame and cm-frame
are easily related:
\beq
{\bf q}_{lab} = {\bf q}_{cm} \, \frac{\sqrt s}{m_N} \qquad .
\eeq
For the decay width $\Gamma$ one finds $\Gamma \sim f^2\,{\bf q}^2$. 
Since the coupling constant $f$ is determined from the value of $\Gamma$
for an on-shell resonance (see Eq. \ref{coupldet}), 
one has (in obvious notation):
\beq
\Gamma \sim f_{lab}^2\,{\bf q}_{lab}^2 = f_{cm}^2\,{\bf q}_{cm}^2 \quad,
\eeq
thus leading to the relation
\beq
f_{lab}^2 = f_{cm}^2\,\frac{m_N^2}{m_R^2} \qquad .
\eeq
Similarly, the forward scattering amplitude is proportional to $f^2\,{\bf q}^2$.
The most relevant contribution from each resonance to the self energy 
comes from the kinematical region $\sqrt s \approx m_R$, where
\beq 
T_{lab} = T_{cm} \equiv T \quad,
\eeq 
as long as lab-frame or cm-frame kinematics are used consistently. However, 
applying mixed kinematics as in \cite{fp,pp98,rw,rw2}, where the coupling
constant is determined within cm-kinematics and the forward scattering amplitude
in lab-frame kinematics, produces 
\beq
T_{mix} = f_{cm}^2\,{\bf q}_{lab}^2 = T\,\frac{m_R^2}{m_N^2} \qquad .
\eeq
Typically, the resonance masses vary between $1.5$ and $2$ GeV. Thus 
for $p$-wave resonances the ambiguity in the non-relativistic reduction
leads to uncertainties which can be as large as a factor of four.
For $s$-wave resonances the effects are less pronounced. Here the transverse
and the longitudinal channel are proportional to the energy 
$\omega^2$ and the invariant mass $q^2$ of the $\rho$ meson, respectively. 
Thus only the transverse channel will be affected by the ambiguity
of the reference frame. Noticeable uncertainties for the $s$-wave part 
of $\Sigma_{med}$
are expected for three-momenta comparable to the invariant mass of the $\rho$.
Altogether, this is clearly not a satisfying 
situation which can only be resolved within a relativistic approach.


\section{Relativistic Framework}
\label{rel}

In this Section we set up the relativistic framework for the calculation
of $\Sigma_{med}$. We will present the Lagrangians used to describe
the $RN\rho$ interaction. A discussion of the resonance propagator
will be deferred to Sect. \ref{resprop}.

The vertex functions ${\cal V}^\mu_\frac12$ and ${\cal V}^{\mu\nu}_\frac32$
are derived from the following relativistic and gauge invariant Lagrangians:
\beq
\label{lagrangian}
{\cal L}_{int} = \left \{
\begin{array}{cclll}
\frac{f_{RN\rho}}{m_\rho}\,{\bar \psi}^{\mu} \, \gamma^\nu \, \psi 
\, F_{\mu \nu} + h.c.  & \quad \mbox{for} & J^\pi = \frac{3}{2}^-\\ \\
\frac{f_{RN\rho}}{m_\rho}\,{\bar \psi}^{\mu} \, \gamma^5 \, \gamma^\nu \, \psi 
\, F_{\mu \nu} + h.c. & \quad \mbox{for} &  J^\pi = \frac{3}{2}^+\\ \\
\frac{f_{RN\rho}}{m_\rho}\,{\bar \psi} \, \gamma^5\,\sigma^{\mu\nu} \, \psi 
\, F_{\mu \nu}+ h.c. & \quad \mbox{for} & J^\pi = \frac{1}{2}^-\\ \\
\frac{f_{RN\rho}}{m_\rho}{\bar \psi}\,\sigma^{\mu\nu} \, \psi 
\, F_{\mu \nu} + h.c.& \quad \mbox{for} & J^\pi = \frac{1}{2}^+ \qquad .
\end{array} \right .
\eeq
Here $\psi^\mu$ denotes the resonance spinor and $\psi$ the nucleon spinor,
$\sigma^{\mu\nu} = \frac{i}{2}\,\left[ \gamma^\mu , \gamma^\nu \right]$ and
$F^{\mu \nu} = \partial^\mu \, \rho^\nu -  \partial^\nu \, \rho^\mu$. 
We note in passing that from these Lagrangians the non-relativistic expressions from  
our previous publication \cite{pp98} are readily obtained. 
We require gauge invariant interactions in order to ensure
physical results in the vicinity of the photon point $q^2= 0$.

The coupling of spin-$1$ particles to the baryon resonances is not uniquely
determined by its quantum numbers. 
In analogy to the commonly used couplings for the $\Delta\,N\,\gamma$ system, we 
employ -- in addition to the couplings in Eq. \ref{lagrangian} --
the following couplings in the $p$-wave channel (see for comparison e.g.
\cite{feu,mukho}):
\beq
\label{coupling}
{\cal L}_{int} = \left \{
\begin{array}{l}
\frac{f_{RN\rho}}{m_\rho^2}
\, {\bar \psi}^\mu \,\gamma^5\,\partial^\nu \psi \, F_{\mu\nu} + h.c. 
\\ \\  \frac{f_{RN\rho}}{m_\rho^2}
\, {\bar \psi}^\mu \,\gamma^5\,\psi \, \partial^\nu F_{\mu\nu} + h.c. \qquad .
\end{array}  \right .
\eeq

In general, the dimensionless coupling constant $f_{RN\rho}$ is different 
in all three cases.
For resonances with negative parity, the corresponding Lagrangians 
${\cal L}_{int}$ can be obtained by dropping the $\gamma^5$-matrix.
In Sect. \ref{uncertain} we will explore the sensitivity of the resonance
model on the chosen vertex function. For the comparison between non-relativistic
and relativistic calculation in Sect. \ref{results}
the coupling from Eq. \ref{lagrangian} will be employed.

In view of the difficulties in constructing a 
relativistic theory for spin-$\frac{5}{2}$ fields, we treat them
non-relativistically. In the relevant kinematical region only
resonances with positive parity couple sufficiently to the $\rho\,N$ system. We 
can therefore restrict ourselves to the discussion of $p$-wave channels.
In agreement with previous studies \cite{fp,pp98} 
the following Lagrangian is used:
\beq
{\cal L}_{int} = \frac{f_{RN\rho}}{m_\rho}\,\chi^\dagger_R\,R^{ij}\,F_{ij}\,\chi  
+ h.c.\qquad .
\eeq
Here $\chi^\dagger_R$ and $\chi$ are resonance and nucleon spinor. $R^{ij}$ denotes
the spin-$\frac{5}{2}$ transition operator \cite{fp}. We will come back
to the treatment of spin-$\frac{5}{2}$ particles in Sect. \ref{results}
and Sect. \ref{uncertain}.

For the isospin coupling we take the standard form:
\beq
\label{iso}
\begin{array}{ccll}
 & \chi_{R}^\dagger\,\sigma_k\,\rho_k\,\chi &\mbox{for}&
 I=\frac{1}{2} \\  \\
 &      \chi_{R}^\dagger\,S_k\,\rho_k\,\chi
                &\mbox{for}& I=\frac{3}{2}  \qquad .
\end{array}
\eeq
$\chi_R$ and $\chi$ denote resonance and nucleon isospinor,
respectively, $S_k$ is the isospin-$\frac{3}{2}$ transition operator and $\sigma_k$ 
the standard Pauli matrix.
 
In the actual calculation each vertex is multiplied by a phenomenological formfactor
\cite{feu}:
\beq
\label{formfac}
F(s) = \frac{\Lambda^4}{\Lambda^4+(s-m_R^2)^2} \qquad .
\eeq
This formfactor has a very different 
functional dependence on $s$ compared to the one previously used in the 
non-relativistic treatment:
\beq
\label{formfacold}
F({\bf q}_{lab}) = \sqrt{\frac{\Lambda^2}{\Lambda^2 + {\bf q}_{lab}^2}} \qquad .
\eeq
We reject this formfactor because it is not Lorentz invariant. Furthermore,
since the resonance excitation is a typical $s$-channel process, the formfactor
should have a maximum on the resonance mass-shell and should only be a function
of the kinematical variable $s$. 
Both criteria are obviously not fulfilled in Eq. \ref{formfacold} since
the three-momentum is a function of the invariant mass $q^2$ of the $\rho$ meson:
\beq
{\bf q}_{lab}^2=\frac{(s-m_N^2-q^2)^2}{4\,m_N^2} - q^2 \quad.
\eeq
For the same reason a formfactor depending on the
cm-momentum of the $\rho$ meson like
\beq
F({\bf q}_{cm}) = \sqrt{\frac{\Lambda^2 + {\bf q}_{0\,cm}^2}
{\Lambda^2 + {\bf q}_{cm}^2}} \quad.
\eeq
is also not satisfactory as ${\bf q}_{cm}$ and ${\bf q}_{0\,cm}$
depend both on $q^2$.
Here ${\bf q}_{0\,cm}$ denotes the $\rho$ momentum in the cm-frame if the decaying 
resonance is on its mass-shell. 

For the cutoff parameter we take a value of $\Lambda = 1$ GeV for all resonances,
in accordance with \cite{feu}. 
The influence of the new formfactor Eq. \ref{formfac}
on the results will be discussed in Sect. \ref{results}.


\subsection{The Resonance Propagator}
\label{resprop}

In this Section we discuss the propagator of the formed baryon resonance.
A proper description of this quantity is not a 
trivial task since it displays in general -- already in the spin-$\frac12$ sector --
a very rich spinor structure (see e.q. \cite{henning}). 
We show that a commonly used recipe for both the propagators of spin-$\frac12$ 
and spin-$\frac32$  particles \cite{adel,dfermion} 
leads to unphysical results in the case of
$\rho\,N$ scattering. Then
we derive an improved 
propagator which -- albeit being only a crude approximation of the 
exact solution -- is nonetheless sufficient for our purposes.

Unitarity relates the matrix element for the decay of 
a resonance $R$ into a nucleon and a $\rho$ meson of given polarization
$|{\cal M}_{RN\rho}^{T/L}|^2$ to the imaginary part 
of the $\rho\,N$ forward scattering amplitude $T^{T/L}$:
\beqn
\mbox{Im } T^{T/L} &\sim& \sigma_{\rho\,N}^{tot} \nonumber \\
           &\sim& \Gamma_\rho\,\Gamma_{tot} \\ 
           &\sim& |{\cal M}_{RN\rho}^{T/L}|^2 \nonumber  \qquad.
\eeqn
It may be objected that for tree-level calculations
unitarity is by no means guaranteed \cite{ben}. 
However by considering only the contribution of one resonance in the
$s$-channel, unitarity can be fulfilled provided 
that the decay width and and the resonance propagator are parameterized in
a consistent way. 

Let us start with the calculation of the squared matrix element for the resonance 
decay  $|{\cal M}_{RN\rho}^{T/L}|^2$. The standard evaluation of this 
quantity, that contains a trace over the internal degrees of freedom, 
involves the completeness
relation of the resonance spinor. Taking it naively as 
\beq
\label{compl_naive}
\sum_r u_r^R(k)\,{\bar u}_r^R(k) = \slash\hspace{-0.2cm}k+ m_R  
\eeq
even for resonances off the mass-shell, i. e. with  $\sqrt{k^2} \ne m_R$,
leads to negative values for the squared matrix element for
positive parity resonances, which is obviously wrong.
The calculation leading to this results is given in the Appendix. 
Thus, Eq. \ref{compl_naive} is not appropriate in the off-shell case and needs
to be modified.
The obvious choice is to set \cite{feu2,shyam} 
\beq
\label{complete}
\sum_r u_r^R(k)\,{\bar u}_r^R(k) = \slash\hspace{-0.2cm}k+\sqrt{k^2} \quad . 
\eeq

Unitarity now enforces that the modified completeness relation Eq. \ref{complete}
also changes the functional dependence of the (imaginary part of the)
propagator. For a spin-$\frac12$ particle one thus obtains
\beq
\label{better}
D_{\frac{1}{2}}(k) = 
\frac{\slash\hspace{-0.2cm}k + \sqrt{k^2} }{k^2-m_R^2+i\sqrt{k^2}\,\Gamma} 
\eeq
instead of 
\beq
\label{naive}
D_{\frac{1}{2}}^{trial}(k) = 
\frac{\slash \hspace{-0.2cm}k + m_R}{k^2-m_R^2+i\sqrt{k^2}\,\Gamma} \qquad .
\eeq
The latter form is often employed in the
literature \cite{adel,dfermion}. Since -- from unitarity -- it is related
to the completeness relation with $m_R$, 
it leads to negative values for the total $\rho\,N$ cross-section
$\sigma_{\rho\,N}^{tot}$.

In the Appendix we give some theoretical backing for Eq. \ref{better}.
In particular, we show that Eq. \ref{naive} violates a fundamental
constraint for the propagator and should therefore be rejected.
We also 
discuss that Eq. \ref{better} is only correct for the imaginary part of 
$D_{\frac{1}{2}}(k)$,
but not for the real part. In an explicit calculation we have checked however,
that the differences between the exact solution and $\mbox{Re}\,D_{\frac12}(k)$
as taken from Eq. \ref{better} are negligible.

The propagator $D_\frac{3}{2}^{\alpha\beta}(k)$
of non-interacting spin-$\frac32$ particles 
is given by the well known  Rarita-Schwinger propagator \cite{rs}. 
The conventional choice of adding a finite decay width in the denominator 
\cite{dfermion,pasca} leads -- as in the 
case of spin-$\frac12$ particles -- to negative total cross-sections for 
processes including intermediate $J^\pi = \frac32^+$ resonances. 
As explained in the
Appendix we propose therefore that the propagator should read:
\beqn
\label{32better}
D_\frac{3}{2}^{\alpha\beta}(k) &=& \frac{\slash\hspace{-.2cm}k + \sqrt{k^2}}
{k^2-m_R^2+i\,\sqrt{k^2}\,\Gamma} \, \\ && \times
\left[ g^{\alpha\beta} - \frac{1}{3}\,\gamma^\alpha \, \gamma^\beta
- \frac{2}{3}\,\frac{k^\alpha\,k^\beta}{k^2} + 
\frac{1}{3}\,\frac{k^\alpha\,\gamma^\beta-k^\beta\,\gamma^\alpha}{\sqrt{k^2}}
\right] \qquad . \nonumber
\eeqn

Some remarks about this propagator are in order.
First we note that it is directly proportional to the 
spin-$\frac{3}{2}$ projector $P_\frac32^{\alpha\beta}$, see \cite{ben}.
Second, in analogy to the case of spin-$\frac12$ resonances this expression 
is -- strictly speaking -- not correct for
the real part, but again we expect the deviations to be negligible
for our purposes.
Finally, Eq. \ref{32better} has already been discussed in the literature
\cite{ben,adel,dfermion}. In \cite{adel} the replacement
$m_R \rightarrow \sqrt{k^2}$ is introduced to insure gauge invariance.
This is not necessary in our case as 
the vector mesons are introduced via the gauge invariant
field tensor $F^{\mu\nu}$. On the other hand, in \cite{ben} it was shown
that Eq. \ref{32better} is incorrect in the limit of a small 
decay width $\Gamma$. 
This observation is in full agreement with our previous statements
about the real part of Eqs. \ref{better} and \ref{32better}.
Also, as pointed out in \cite{ben,dfermion}, the propagator Eq. \ref{32better} 
displays a pole at $k^2 = 0$, which can lead to divergent results for scattering 
amplitudes.
However, this kinematical regime is reached only
in $u$-channel processes which we do not include in our model. 
Therefore the results of this work should not suffer from these poles.
One should also be aware of the 
fact that the use of gauge-invariant couplings for the spin-$\frac32$ field --
which introduce additional orders of the  resonance momentum --
as suggested in \cite{pasca} avoids these poles.



\section{Included Resonances And Their  Parameters}
\label{manley}

In this Section we list the resonances included in our model for 
$\Sigma_{med}$ and discuss their parameters such as mass, decay width and branching ratios.

We include all baryon resonances with a sizeable coupling
to the $\rho\,N$ channel. 
They are listed in Table \ref{barres}.
The nucleon is no longer included in the model. We found in a relativistic calculation
on the $T$-matrix level
that the Born terms lead to a strong overestimation of the data for the reactions
$\gamma\,N \rightarrow \rho\,N$ and $\pi\,N \rightarrow \rho\,N$, as well as for reactions
with an $\omega$ meson in the final state. It is very likely, that on this level important 
backscattering effects are missing, since the nucleon has a large coupling strength
to the $N\,\pi$, $N\,\omega$ and $N\,\rho$ channels. Similar effects are known
from Compton scattering, where in a $K$-matrix analysis,
which includes the rescattering, the Born terms
yield satisfactory results. In a  $T$-matrix approach on the other hand,
additional formfactors, which suppress the contribution in backward direction, have to
be introduced in an {\it ad-hoc} manner \cite{feu3}. 
Therefore it does not seem appropriate to include the nucleon in our model.
For resonances, on the other hand, 
backscattering effects are effectively taken into account by their decay width.

In contrast to our previous publication \cite{pp98}, we take the resonance 
parameters from the 
analysis of Manley {\it et al} \cite{man1,man2} rather than those 
from the PDG group \cite{pdg}. The PDG parameters represent estimates 
of resonance properties, averaging the available experimental results. 
They are based on a variety of theoretical models, which in general
differ in many details, as, for example, the included resonances 
or the treatment of the non-resonant background.
Therefore, the extracted parameters are strongly model 
dependent and it is questionable whether an average of the results
from different studies provides conclusive information on any specific reaction channel.
To ensure a realistic description of the $\rho\,N$ scattering amplitude,
it is much safer to stick to the results of a single analysis
of $\pi\,N \to \pi\,\pi\,N$ reactions.
In the work of Manley {\it et al} the available data set for
the reaction $\pi\,N \rightarrow \pi\,\pi\,N$ was analyzed within an isobar model,
which allows for various intermediate $\pi\,\pi\,N$ states, such as $\Delta \,\pi$
and $N\,\rho$. All partial waves were fitted with resonance plus background
terms, and from that fit the resonance parameters were extracted. 
Similar work has been done by various other groups \cite{long}. 
We follow the results of Manley {\it et al}, 
since his analysis is the most extensive and recent one. 
In this way, rather than by
using the parameters of \cite{pdg}, experimental information on 
$\rho\,N$ scattering enters into the model in a well defined way.

In Table \ref{barres} we list the mass $m_R$, the total decay width $\Gamma_{tot}$ and
the decay width into the $\rho\,N$ channel, as taken from \cite{man2}.
For comparison, we also give the value for the $\rho\,N$ decay width from
\cite{pdg}. In the last column we denote if the respective resonance
decays in a relative $s$-wave or $p$-wave into the $\rho\,N$ channel.

A considerable uncertainty in the model is connected 
with the coupling strength of some
resonances to the $\rho\,N$ channel. This is most evident for transversely 
polarized $\rho$ mesons. Here, at intermediate and high momenta, 
the $P_{13}(1720)$ resonance plays a dominant role, 
due to its large partial decay width into the $\rho\,N$ channel as given
in the analysis of Manley {\it et al} \cite{man2}.
As listed in Table \ref{barres}, the estimate of the PDG group \cite{pdg}
is smaller by a factor of 3. Moreover, in Manley's analysis a large error of 
$168$ MeV - which is of about the same size as the decay width itself - is stated. 
Similarly, for the $D_{33}(1700)$ the partial decay width as given by PDG and by Manley
differ by a factor of 3, mainly affecting the results of $A_\rho^L$.
Again, Manley gives a very large error of $31$ MeV for this
quantity. On the other hand, very good agreement is reached on the properties
of the $D_{13}(1520)$. Here, Manley and the PDG group present nearly 
identical values for the partial decay width. Furthermore, the error for this number is 
much smaller, only $6$ MeV are listed in \cite{man2}.

In Manley's fit a few high lying resonances, which have not been included in our previous
calculation, show a sizeable branching ratio into the $\rho\,N$ channel.
They are listed below the double-line in Table \ref{barres}. However, as 
will be discussed in 
Sect. \ref{results}, they induce only small changes on the results. 

The total decay width $\Gamma$ of the resonances is taken as a sum over
all partial decay channels. The branching ratios into these channels as well
as the mass dependence of the decays are the same as in \cite{man2}, except for 
the $\rho\,N$ decay. This decay channel is calculated within the same framework 
as the forward scattering amplitude, see Eq. \ref{coupldet}.

Due to the small phase space there is no measured decay width of the $P_{33}(1232)$
into $\rho\,N$. Therefore the coupling constant of this resonance to the $\rho\,N$ system
can not be obtained by a fit to the decay width
and is subject to substantial uncertainty.
In our previous calculation, as well as in other works \cite{rw,rw2}, it was taken 
from the Bonn potential model. It is however questionable, whether this value
has much meaning outside the Bonn model (cf. discussion in \cite{fl}). 
We therefore calculate
the coupling constant directly from the well known isovector part of the 
$\Delta \rightarrow N\,\gamma$ decay width, assuming the validity of strict VMD \cite{pm}.
Note that the numerical values obtained in both ways are nearly identical.
The mass dependence of the decay width is obtained in the same way as for the other
resonances.

It might be argued, that the consistency of the resonance parameters employed for 
the calculation of the $\rho\,N$ forward scattering amplitude can be tested by
comparing with data on the reaction $\pi\,N \rightarrow \rho\,N$ \cite{brody}.
On inspection of these data, there seems to be 
a lack of strength below the threshold at roughly $1.7$ GeV, in conflict with a
large coupling of the $D_{13}(1520)$.
However, this conclusion is not correct. The $\rho\,N$ ``data points''
were extracted from the measured $\pi\,\pi\,N$ data by simple fits of the mass differential
cross section. Naturally, with such a procedure no sensitivity is achieved in the 
subthreshold region, where the $\rho$ meson can not directly be identified from the 
$\pi\,\pi$ mass spectrum. Such a fit
is certainly much less conclusive than the partial wave analysis 
of Manley {\it et al}. Moreover,
the data set used by Manley {\it et al} also comprises the 
$\pi\,N \rightarrow \pi\,\pi\,N$ data from \cite{brody}.


\section{Results 1 - Relativistic Versus \\  \hspace{14cm} Non-Relativistic}
\label{results}

In Sect. \ref{review} we discussed the shortcomings of the non-relativistic
calculation of the self energy $\Sigma_{med}^{T/L}$ of the $\rho$ meson. We then 
developed the framework for a fully relativistic calculation. Now we compare
the non-relativistic and the relativistic approach -- using the relativistic
Lagrangians from Eq. \ref{lagrangian} --
 and pay special attention to
the question, whether a reference frame exists, in which the non-relativistic
reduction should be performed in order to arrive at a good approximation
of the relativistic results.

\begin{figure} 
\begin{center}
\epsfig{file=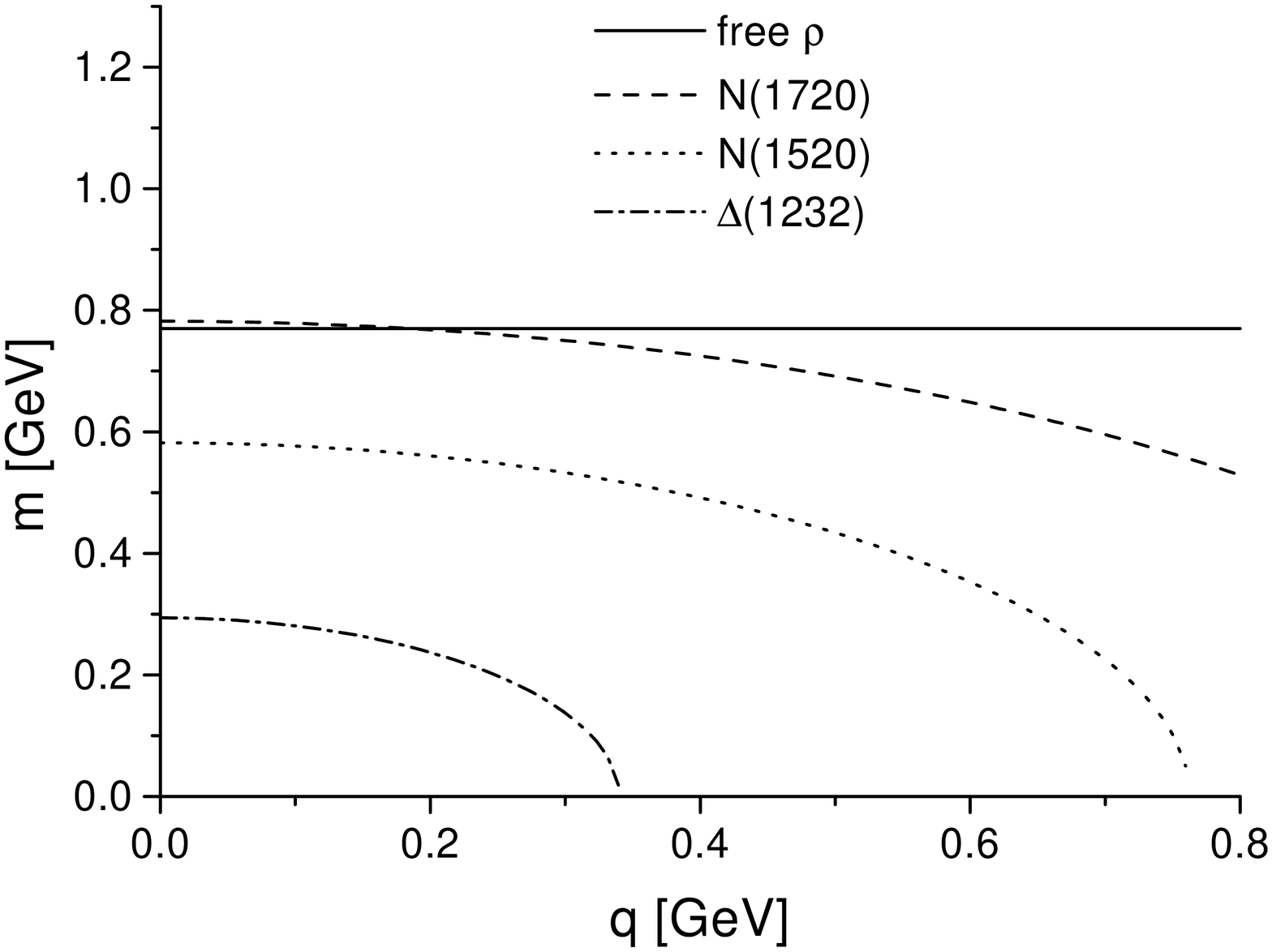,width=10cm}
 \caption{ 
\label{branches}
Free dispersion relations for the $\rho$ and the most prominent resonances.
For the resonance branches $m$ is the invariant mass a meson of given 
three-momentum ${\bf q}$ must have in order to excite the respective resonance on
its mass shell \cite{pp98}.}
\end{center}
\end{figure}

In a resonance model, neglecting level repulsion for the moment, 
the spectral function $A_\rho$ is 
most sensitive to the behaviour of the scattering amplitude 
in the vicinity of $s = m_R^2$, i.e. along the lines depicted in Fig. ~\ref{branches}.  
In order to get a feeling for the differences between a relativistic 
and the different non-relativistic descriptions,
it is therefore rewarding to compare the results from both approaches
for the quantity $\Omega^{T/L}$, which relates to the forward scattering
amplitude as:
\beq
\label{spur}
T^{T/L} = - \Sigma_I\,F(k^2)^2
\frac{\Omega^{T/L}}{k^2-m_R^2 + i\,\sqrt {k^2}\,\Gamma} \quad.
\eeq
Here $k^2$ stands for the invariant mass of the $\rho\,N$ system;
the formfactor $F$ and the isospin factor $\Sigma_I$ were already introduced 
in Sect. \ref{rel}.
Note that the coupling constants $f_{RN\rho}$ are included in the definition
of $\Omega^{T/L}$ and have been 
adjusted to the partial decay width of the resonance.
The quantities $\Omega^{T/L}$ are explictly given for both a non-relativistic
and a relativistic calculation in the Appendix.

In Fig. \ref{spurt} we show
$\Omega^T$ for two of the most prominent resonances in the 
non-relativistic analysis, the $D_{13}(1520)$ and the $P_{13}(1720)$.
It is plotted as a function of the invariant mass of the $\rho$ meson, the momentum
being fixed by the condition $s = m_R^2$. We compare the relativistic results
with three non-relativistic versions of $\Sigma_{med}^{T/L}$, which are 
explained in detail in Sect. \ref{review}:
\begin{itemize}
\item[i)] relativistic 
\item[ii)] non-relativistic, selfenergy lab-frame, coupling constant cm-frame
\item[iii)] non-relativistic, both quantities cm-frame
\item[iv)] non-relativistic, both quantities lab-frame
\qquad . 
\end{itemize}
Up to now, in the literature only calculations
performed within kinematics ii) can be found \cite{fp,pp98,rw,rw2}.

We first remark that the results are not degenerate at threshold, since   
in each case the coupling constant has been adjusted to the measured partial
decay width. As expected from the discussion in Sect. \ref{review},
the calculations reveal large differences between 
the different non-relativistic approaches ii)-iv) in the transverse channel.
This effect is -- as already discussed -- more pronounced for $p$-wave 
resonances, see lower graph of Fig. \ref{spurt}. Here,
especially at low invariant masses, the commonly used
non-relativistic version ii) is not reliable at all, 
overestimating the results by a factor of $2-3$, whereas both version iii) and iv) 
yield a surprisingly good agreement with the fully relativistic calculation i).
In the $s$-wave channel -- upper graph of Fig. \ref{spurt} -- it is again
the non-relativistic version ii) displaying a considerable disagreement with
the relativistic calculation. Both versions iii) and iv) yield better approximations.
The exact form of the momentum dependence, however, is not well reproduced in 
version iv). We note that also in \cite{koch2} a good agreement between a relativistic 
calculation for the dilepton decay of the $D_{13}(1520)$ 
and a non-relativistic one performed in the cm-frame was reported.

We do not show results for $\Omega^L$. Here
the effect from a relativistic calculation is rather small, 
since the momentum dependence
of $\Omega^L$ is mainly determined
from physical constraints at ${\bf q}=0$ and at the photon point:
at small momenta a relativistic calculation and its non-relativistic reduction must 
display a similar behaviour, whilst at the photon point
$\Omega^L$ has to vanish in either version.
Thus, both calculations do not have much room to develop differently.
There is a finite contribution to $\Omega^L$ from $p$-wave resonances, which, however,
is small as compared to the transverse coupling.
In the spin-$\frac{1}{2}$ sector, one obtains qualitatively similar results for 
$\Omega^{T/L}$.

Summarizing these results, we find that the previously used non-relativistic version
ii) is not a good approximation of a fully relativistic calculation. In contrast,
using cm-kinematics for both the determination of the coupling constant and the
self energy leads to a very reasonable agreement with the relativistic approach.

For this reason we argue also that 
the contribution from spin-$\frac{5}{2}$ particles
--- for which a relativistic theory is very complicated --- 
can be reliably evaluated in the non-relativistic approach iii).
Since in a non-relativistic calculation the vertex factors of both the spin-$\frac{3}{2}$
and spin-$\frac{5}{2}$ resonances are proportional to ${\bf q}^2$, we expect from
Fig. \ref{spurt} that at high momenta the contribution from spin-$\frac{5}{2}$
resonances is reduced as compared to previous works \cite{fp,pp98,rw}, where
${\bf q}^2$ was evaluated in the lab-frame.

\begin{figure}
\begin{center} 
\epsfig{file=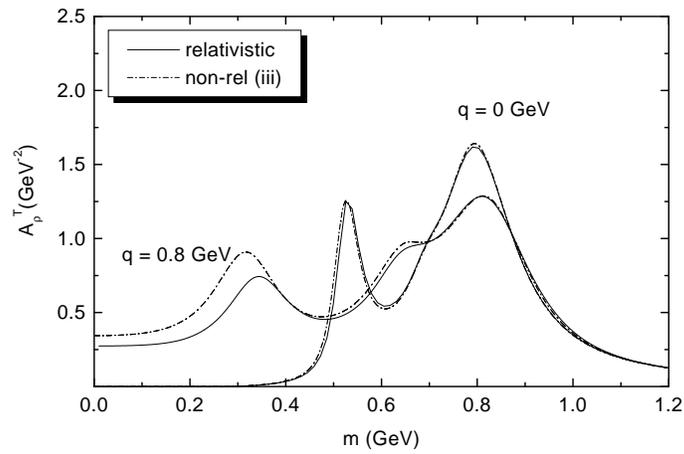,width=10cm}
 \caption{ 
\label{relkin3} The transverse spectral function $A_\rho^T$ at $0$ GeV and
$0.8$ GeV for $\rho=\rho_0$ as a function of the invariant mass $m$ of the 
$\rho$ meson.
Shown is a comparison between a relativistic calculation and
and a non-relativistic one using cm-kinematics iii). }
\end{center}
\end{figure}

Turning now to the discussion of the results for the spectral function, we
start with a comparison between a relativistic calculation and the non-relativistic
approach with kinematics iii). These and all other  
calculations of the spectral function have been 
performed at normal nuclear matter density $\rho_0 = 0.17/{fm^3} $.
The good agreement for the results of $\Omega^{T/L}$  translates into
very similar results for the spectralfunction $A_\rho^{T/L}$, see Fig. \ref{relkin3}.
There the transverse spectralfunction is plotted at momenta $q=0$ GeV and $q=0.8$ GeV.
The resonance parameters are taken from Manley {\it et al}, only the resonances
above the double line in Table \ref{barres} are included. The small deviations
at large momenta are partly due to the different choices for the formfactor, see Eqs.
\ref{formfac} and \ref{formfacold}. The effect of the formfactor will be 
discussed in more detail later on in this Section. We do not present a comparison
for the longitudinal channel as there the differences are even smaller.

\begin{figure}
\begin{center} 
\epsfig{file=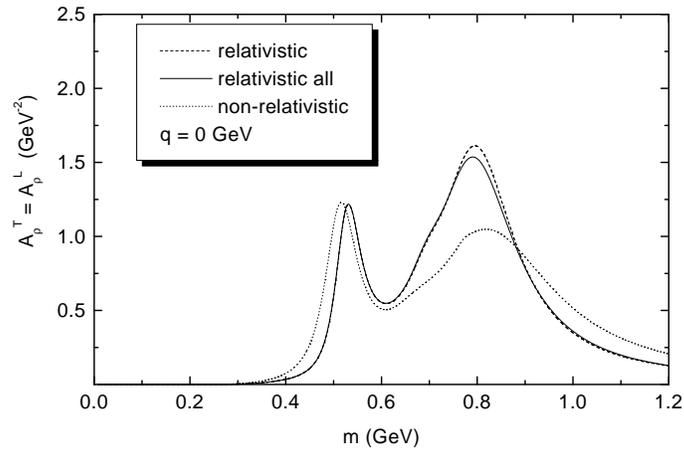,width=10cm}
 \caption{ 
\label{a_rnr0}
$A_\rho^T = A_\rho^L$ at ${\bf q}= 0$ for $\rho=\rho_0$ 
as a function of the invariant mass $m$ of the 
$\rho$ meson. 
Compared are a relativistic (solid line, dashed line)
and a non-relativistic calculation (dotted line). For the relativistic
calculation the resonance parameters of Manley {\it et al} \cite{man2} are employed,
for the non-relativistic one the parameters of PDG \cite{pdg}. }
\end{center}
\end{figure}

Since the agreement between
these two approaches is nearly perfect we will use the remainder of this Section
for the comparison of our previous non-relativistic results (within kinematics ii))
and the relativistic ones. Differences between both versions are not only
due to kinematics. Also the changes in the resonance parameters and the 
influence of the new formfactor need to be considered. This is
apparent already at low momenta, see Fig. \ref{a_rnr0}. 
For the moment we concentrate on a comparison of
our previous non-relativistic
calculation (dotted line) with the relativistic one, 
where the same resonances
as in the non-relativistic case (listed above the double-line in
Table \ref{barres}) are included (dashed line).
At ${\bf q} = 0$ the
non-relativistic calculation with PDG parameters shows a stronger depletion
of the $\rho$ peak than the relativistic approach with Manley's parameters.

The formfactor $F(s)$ in Eq. \ref{formfac} suppresses off-shell 
contributions $(s \ne m_R^2)$ of resonances and thus ensures that
at large invariant masses of the $\rho$ meson the forward scattering 
amplitude becomes negligible. It therefore provides a conservative estimate of 
the resonance contribution.
Such a behaviour is not guaranteed by 
using the old formfacor Eq. \ref{formfacold} instead, which is equal to 
one at ${\bf q}=0$. With this formfactor the forward scattering amplitude
increases with the invariant mass of the $\rho$, due to the energy dependence of
the vertex function $\Omega^{T/L}$.
This affects in general both
the tail of the resonance excitation and, due to level repulsion, the strength
in the resonance peak.
Thus, due to the formfactor, the impact of the excitation of the $D_{13}(1520)$ 
in the region of the $\rho$ peak is noticeably reduced.
Since the formfactor has such an impact on the results, we investigated
the sensitivity on the chosen value for the cutoff parameter $\Lambda$ by
varying $\Lambda$ in the range of $1-1.5$ GeV. The effects on the results
for the spectral function were found to be small, they are at the order of $10\%$.
This effect is enhanced by the small partial $\rho\,N$ decay width 
of the $D_{33}(1700)$ found in the Manley analysis as compared to 
PDG (ct. Table \ref{barres}). Its small coupling
makes this resonance play a merely negligible role in the relativistic calculation.

\begin{figure}
\begin{center} 
\epsfig{file=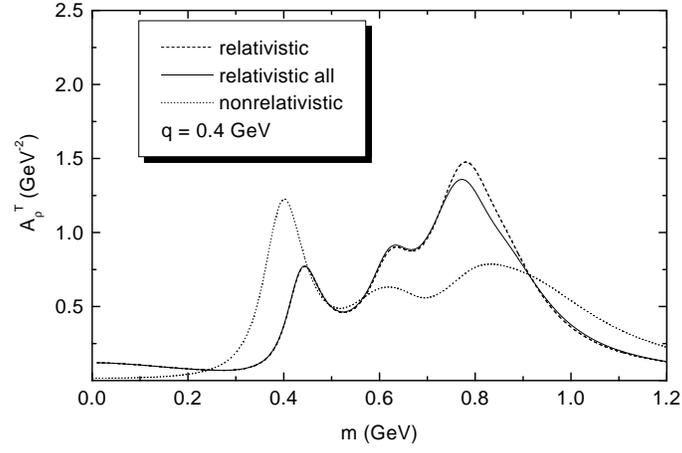,width=10cm}
\epsfig{file=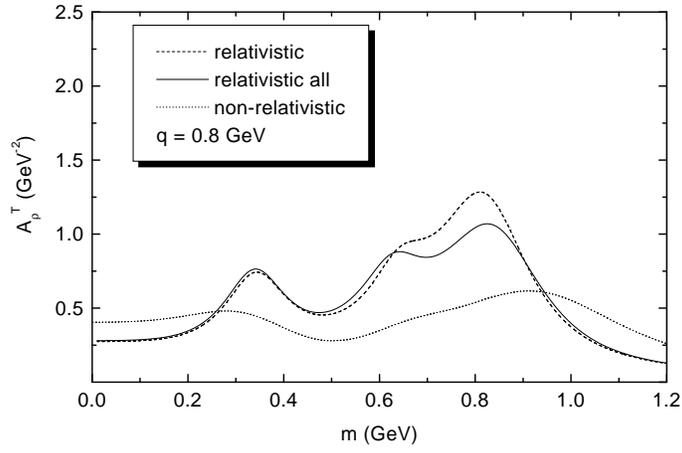,width=10cm}
 \caption{ 
\label{a_rnr4t}
$A_\rho^T$ at ${\bf q}= 0.4$ GeV (upper part) and 
at ${\bf q}= 0.8$ GeV (lower part) at $\rho=\rho0$ 
as a function of the invariant mass $m$ of the $\rho$ meson.}
\end{center}
\end{figure}

At larger momenta both approaches lead to very different results for $A_\rho^T$,
see Fig. \ref{a_rnr4t}.
As a general tendency, in the new calculation
much more strength is centered around the original $\rho$ peak.
This behaviour can be retraced to various sources. 
As follows from the properties of $\Omega^{T/L}$, for a 
given branching ratio, a relativistic calculation produces a much smaller 
$\Sigma_{med}^T$ than a non-relativistic one. 
The resonance contribution is further suppressed by the formfactor $F(s)$,
thus producing a more conservative estimate of the 
resonance contribution. In addition, at higher momenta, $\Sigma_{med}$ is reduced
by the smaller branching ratio of the $D_{33}(1700)$. These depletion mechanisms are
only counteracted by the strongly increased branching ratio of the $P_{13}(1720)$.
Overall, at high momenta $A_\rho^T$ is dominated by the 
$D_{13}(1520)$ and the $P_{13}(1720)$ as well as, to some extent,
by the spin-$\frac{5}{2}$ resonances $F_{15}(1680)$ and $F_{35}(1905)$. 
Still the transverse spectral function receives significant broadening with increasing
3-momentum in a relativistic calculation, but this effect is not as pronounced
as in the non-relativistic case.

\begin{figure}
\begin{center} 
\epsfig{file=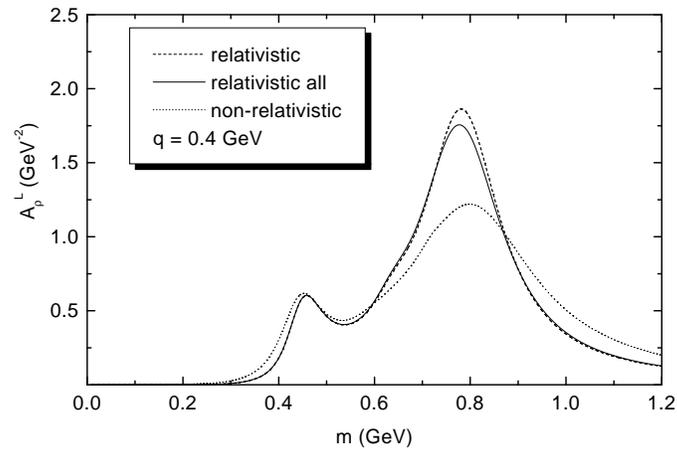,width=10cm}
\epsfig{file=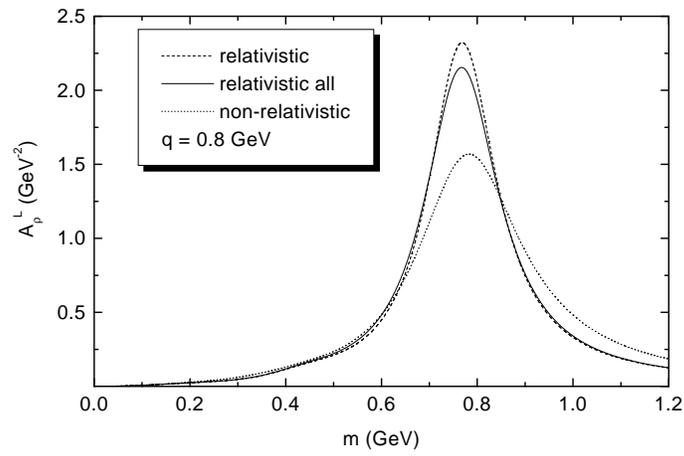,width=10cm}
\caption{
\label{a_rnr4l}
Same as Fig. \ref{a_rnr4t}, but for $A_\rho^L$.}
\end{center}
\end{figure}

Whereas $A^T$ receives strong qualitative modifications in a relativistic 
calculation, changes in $A^L$ are only of a quantitative nature. 
This is shown in Fig. \ref{a_rnr4l}.
As already mentioned, the momentum dependence is dominated --- both 
in a relativistic and a non-relativistic approach --- by physical constraints at
${\bf q} = 0$ and $q^2=0$. Consequently, the $D_{13}(1520)$ peak at ${\bf q}=0.4$ GeV
contains nearly the same strength in both approaches. The depletion 
of resonance strength around
the $\rho$ peak is, as in the case of ${\bf q}=0$, due to the 
formfactor and the small coupling of the 
$D_{33}(1700)$. In the longitudinal channel the only sizeable contribution
comes from the $D_{13}(1520)$.

The impact of the new, high lying resonances on our results is small and noticeable
only at high momenta. This is easily understood for the following reasons:
Despite their large branching ratio, the coupling strength is 
relatively small, since their mass is far above the $\rho\,N$ threshold and consequently
the phase space is very large. In addition, they have a very 
large total decay width, leading to a further suppression of the contribution.
At low momenta, these resonances can only be excited from $\rho$ mesons far
above their pole mass. Correspondingly, their contribution to the imaginary part of 
$\Sigma_{med}$ is drowned by the imaginary part of the vacuum
self energy, the 2 $\pi$ decay width. It is only at momenta around $0.8$ GeV, that
the resonance excitation occurs at invariant masses around the pole mass of the $\rho$
meson. At this point resonance scattering becomes important.
This is shown in Figs. \ref{a_rnr0}-\ref{a_rnr4l}, where the 
calculation including all resonances is represented by the solid line.

\begin{figure}
\begin{center} 
\epsfig{file=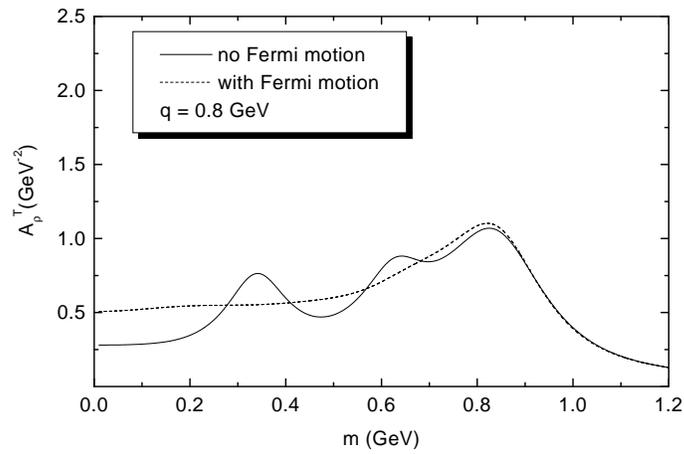,width=10cm}
 \caption{ 
\label{a_med08} $A_\rho^T$ at ${\bf q}= 0.8$ GeV for $\rho=\rho_0$
as a function of the invariant mass $m$ of the 
$\rho$ meson.
The vacuum result (solid line)
is compared to an in-medium calculation where
Fermi averaging and short range correlations are taken into account (dashed line).}
\end{center}
\end{figure}

Finally, we discuss the effect of medium modifications of the scattering amplitude.
The Fermi averaging shows sizeable impact only at large momenta, leading to 
a broadening of the resonance peaks. 
As for the case of Fermi averaging, short range correlations influence the results
mostly at high momenta. We show a comparison of the influence of both modifications
on $A_\rho^T$ at ${\bf q}= 0.8$ GeV in Fig. \ref{a_med08}. For $A_ \rho^L$ the 
differences are completely negligible.


\section{Results 2 - Influence of Couplings}
\label{uncertain}

So far, we have discussed the differences between a relativistic and a non-relativistic
calculation of $\Sigma_{med}$ and their consequences for $A_\rho$. 
However, the Lagrangians from Eq. \ref{lagrangian} is not unique; the
resonance can couple in many more ways to the $\rho\,N$ system.
In order to draw safe conclusions on the momentum dependence of the 
$\rho\,N$ scattering amplitude, it is important to study the influence 
of the chosen coupling on the results.

In this analysis we confine ourselves to the case of spin-$\frac{3}{2}$ resonances
of both positive and negative parity. The vertex functions are derived from
the Lagrangians in Eqs. \ref{lagrangian} and \ref{coupling}.
In the spin-$\frac{1}{2}$ sector there is no
resonance which has a sizeable coupling to $\rho\,N$. Clearly, a strong dependence of 
the final results on the coupling of spin-$\frac{1}{2}$ resonances is not expected.

\begin{figure}
\begin{center} 
\epsfig{file=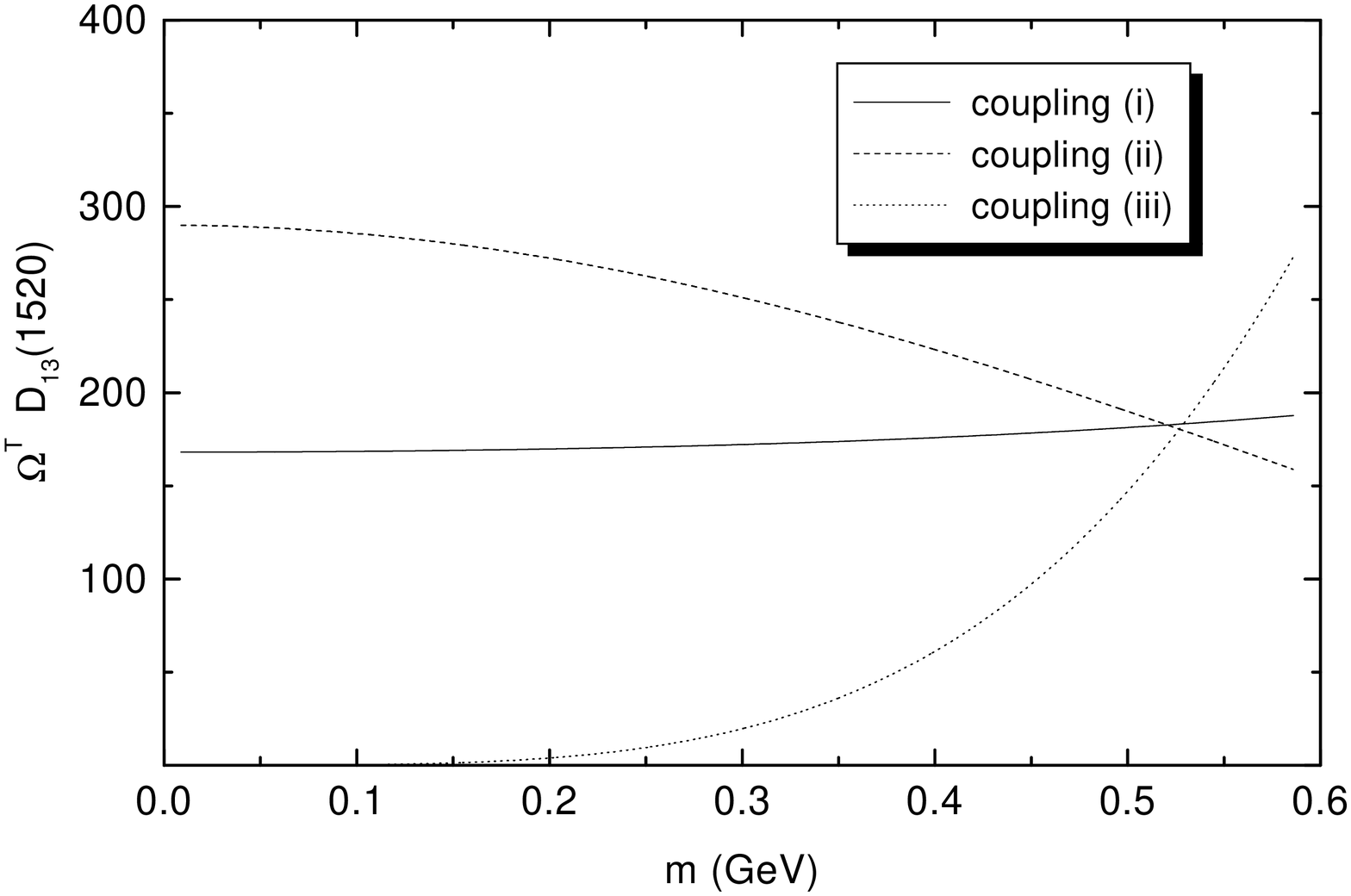,width=10cm}
\epsfig{file=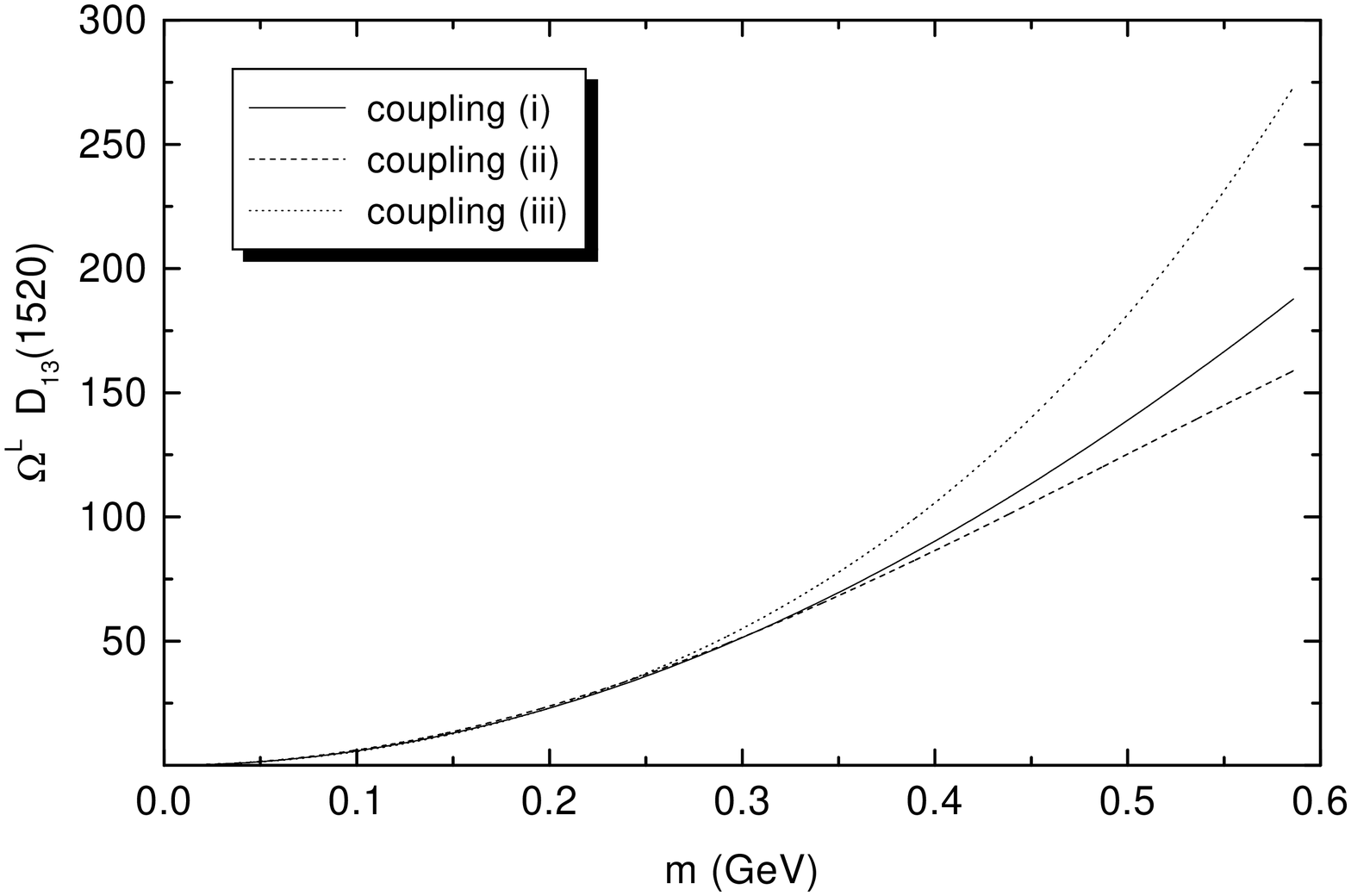,width=10cm}
 \caption{ 
\label{comps} $\Omega^T$ (upper graph) and $\Omega^L$ (lower graph) for 
the $D_{13}(1520)$  at $s=m_R^2$ as a function of the invariant
mass $m$ of the $\rho$ meson.
Shown is the influence of different coupling schemes on the results. The couplings
are explained in Sect. \ref{uncertain}.}
\end{center}
\end{figure}

\begin{figure}
\begin{center} 
\epsfig{file=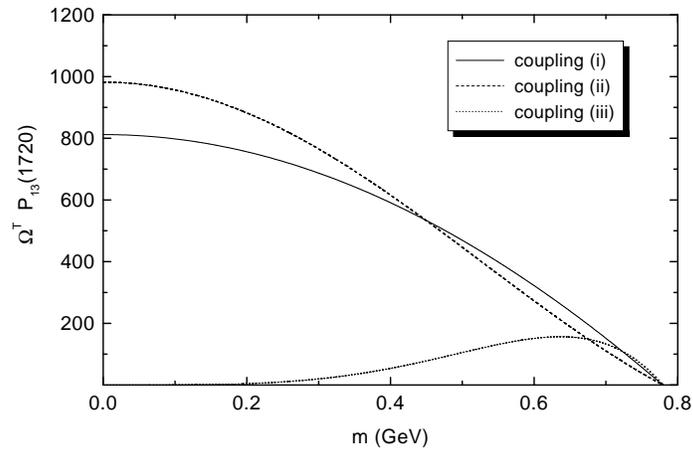,width=10cm}
\epsfig{file=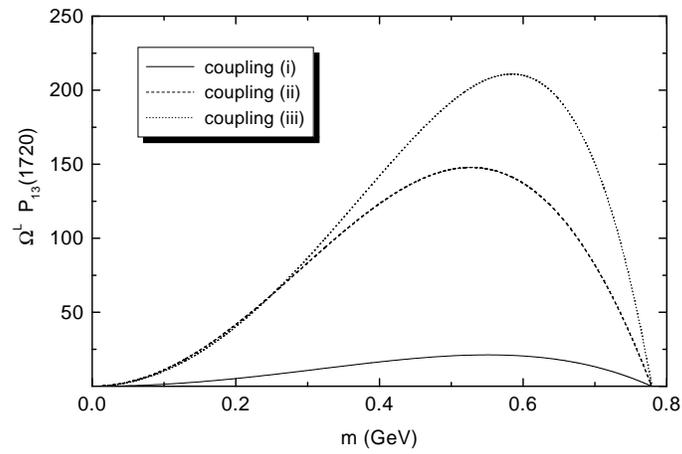,width=10cm}
 \caption{ 
\label{compp} Same as Fig. \ref{comps} but for the $P_{13}(1720)$.}
\end{center}
\end{figure}

In the following discussion we will use some shorthand notations to distinguish
between the different vertex functions. The standard coupling from
Eq. \ref{lagrangian} will be referred to as coupling i), to the first
vertex function from Eq. \ref{coupling} we refer as coupling ii) and
to the second one as coupling iii).
To get a feeling for the changes, we present first $\Omega^{T/L}$ in 
Fig. \ref{comps} and  \ref{compp}. 
The results are shown for the $D_{13}(1520)$ and the $P_{13}(1720)$ resonances.
Explicit results for these quantities are
given in the Appendix. Note that due to dimensional reasons in the definition 
of $\Omega^{T/L}$ in Eq. \ref{spur} an additional factor $\frac{1}{m_\rho}^2$ 
has to be introduced, see all Eq. \ref{coupling}.
The results are nondegenerate at threshold since the coupling constants $f_{RN\rho}$ 
are already adjusted to the $\rho\,N$ 
branching ratios.
The most distinct behaviour is exhibited by coupling iii). At the photon point, 
it enforces a vanishing contribution not only in the longitudinal, but also in the 
transverse channel. This behaviour can be read off from the Lagrangian ${\cal L}_{int}$:

\beq
\partial^\nu \, F_{\mu\nu} = -q^2\,A^\nu
\qquad .
\eeq

Clearly, the vertex function has to vanish at $q^2=0$, regardless of the polarization
of the $\rho$ meson, as can easily be read off the explicit expression 
for $\Omega^{T/L}$ in the Appendix, see Eq. \ref{r3}.
Consequently, $\Sigma_{med}^T$ has to fulfill the same constraints as 
$\Sigma_{med}^L$. Both become more or less identical. Thus, the coupling iii) mostly
affects the transverse channel. Noteworthy is also the much larger contribution 
from $p$-wave resonances to $\Omega^L$. 

For the same $\rho\,N$ branching ratios,
large coupling constants are obtained by using coupling iii), 
as compared to the other
two couplings schemes. In the determination of the coupling
strength the quantity $2\,\Omega^T + \Omega^L$ enters, averaged over the 
$\rho$ mass distribution. Since this expression has to vanish at $q^2 = 0$
in coupling iii), its size is much reduced at low invariant masses.
This behaviour can only be compensated by a large coupling constant.
For $p$-wave resonances this effect is most evident, since in this case
the matrix element becomes sizeable mostly near the photon point for 
the couplings i) and ii).
As an example, we obtain a value of $f_{RN\rho} = 36$ for the $P_{13}(1720)$ 
in coupling iii) as compared to $f_{RN\rho}=10$ in coupling i).
Clearly, the increased values for the coupling constants are 
responsible for the enhanced
resonance contribution in the longitudinal channel.

For coupling iii) the coupling strength for the $P_{33}(1232)$ 
can not be obtained from VMD since
the matrix element vanishes at the photon point. 
We take an {\it ad hoc} value of $f_{RN\rho}=30$ for the coupling constant, which 
is twice as large as that obtained from coupling i). 
Varying $f_{RN\rho}$ between values of $15 - 30$ has only negligible influence on the
results.

Coupling ii) produces much less dramatic changes as compared to iii). 
The general properties of $\Omega^{T/L}$
do not change much in comparison to a calculation employing coupling i).
An exception is the coupling of $p$-wave resonances to longitudinal
$\rho$ mesons, where coupling ii) generates a much larger contribution
than coupling i).
Quantitatively, also the coupling of $s$-wave resonances to transverse $\rho$ 
mesons change up to some extent. 
Nearly identical results are achieved in the remaining channels.

\begin{figure}
\begin{center} 
\epsfig{file=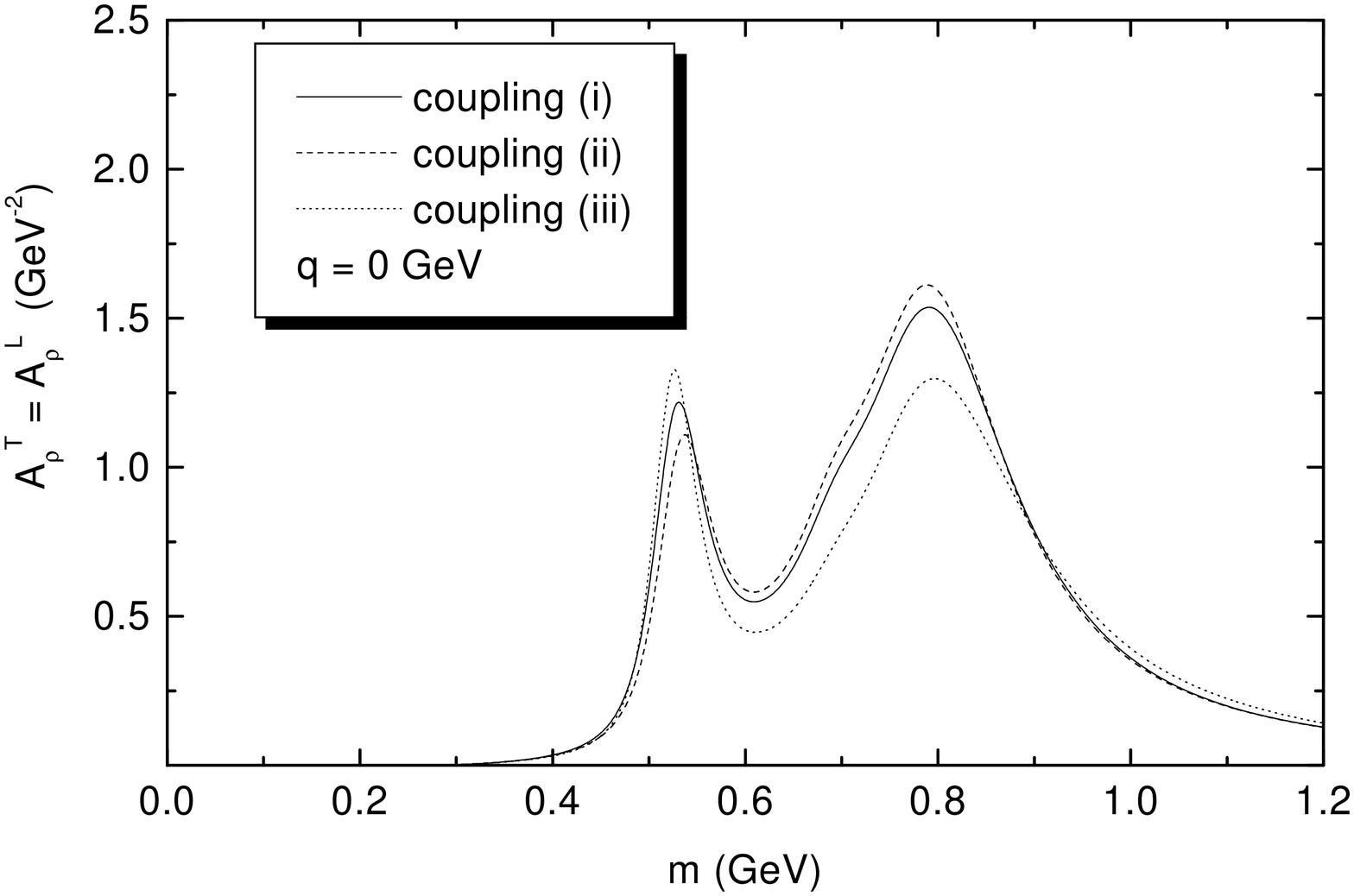,width=10cm}
 \caption{ 
\label{a_comp0}
$A_\rho^T = A_\rho^L$ at ${\bf q = 0}$ GeV for $\rho=\rho_0$
as a function of the invariant mass $m$ of the 
$\rho$ meson.
The couplings employed for the calculations are explained
in Sect. \ref{uncertain}. }
\end{center}
\end{figure}

\begin{figure}
\begin{center} 
\epsfig{file=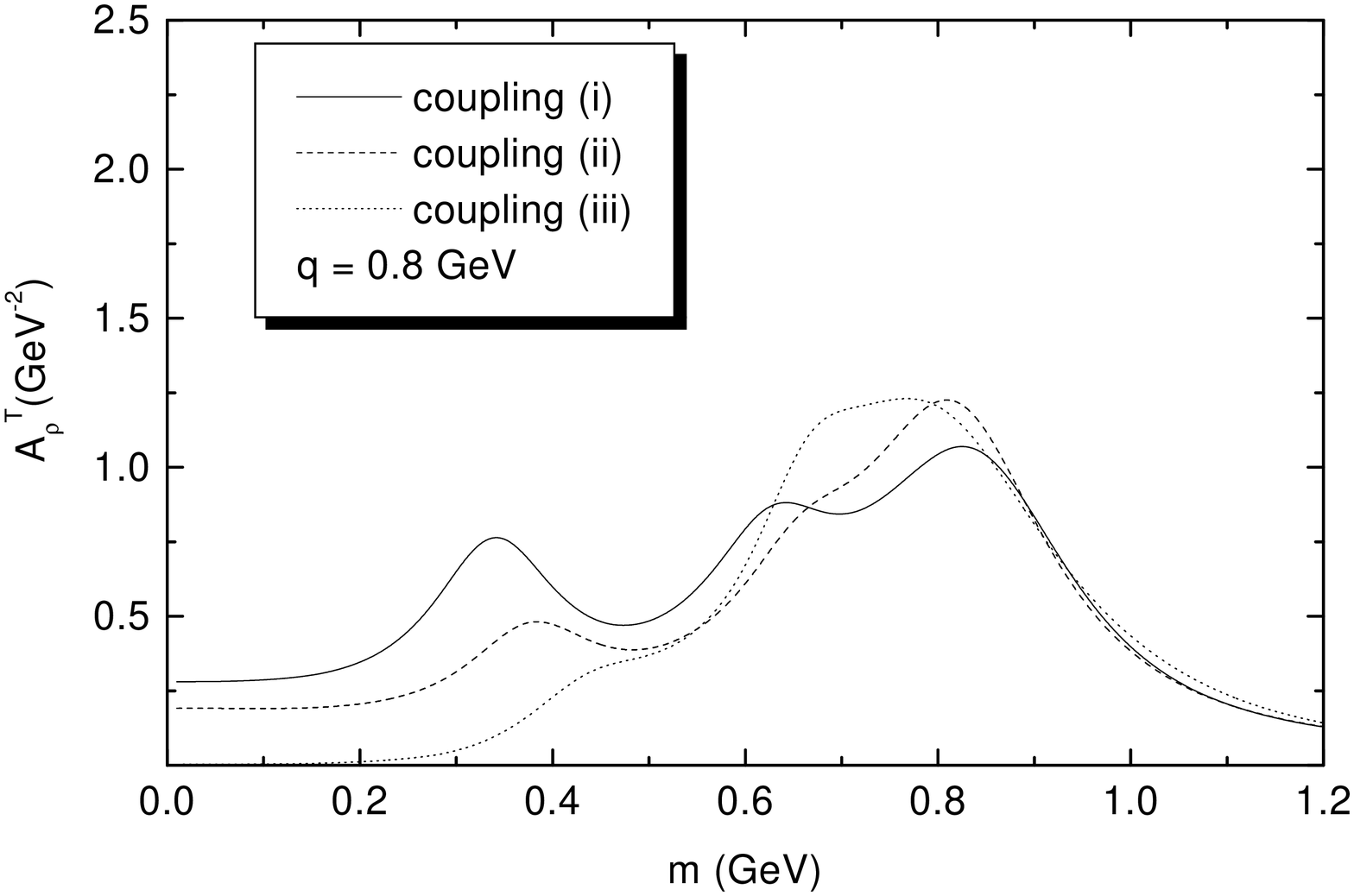,width=10cm}
\epsfig{file=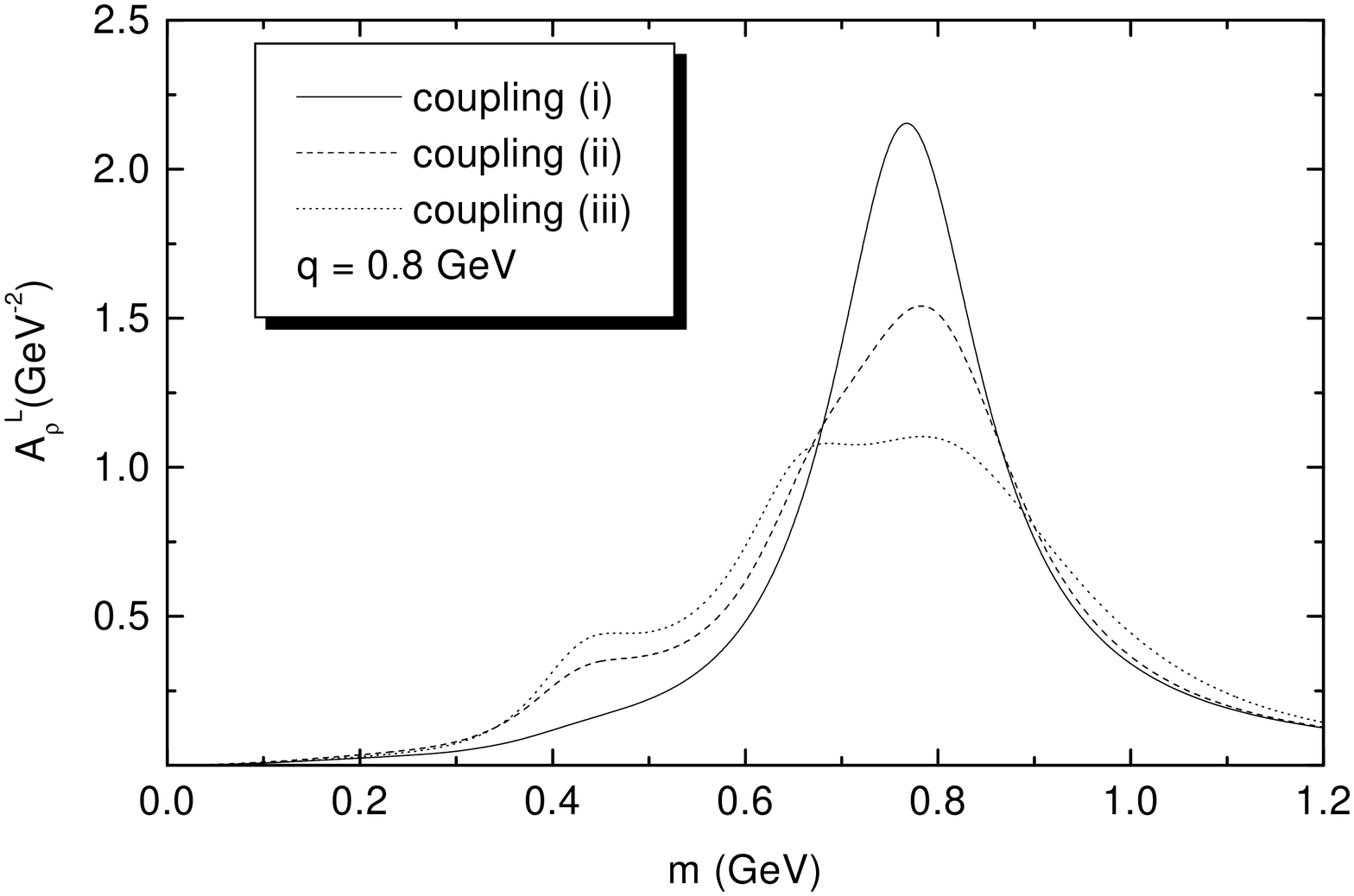,width=10cm}
 \caption{ 
\label{a_comp8t}
$A_\rho^T$ (upper graph) and $A_\rho^L$ (lower graph) at ${\bf q}=0.8$ GeV for 
$\rho=\rho_0$ as a function of the invariant mass $m$ of the 
$\rho$ meson.
The couplings employed for the calculations are explained
in Sect. \ref{uncertain}. }
\end{center}
\end{figure}

Turning now to the results for the spectral function, at ${\bf q}= 0$ no significant 
modification is observed, see Fig. \ref{a_comp0}. However, at higher momenta, the 
distinct character of coupling iii) becomes apparent, reducing the 
resonance contribution at low invariant masses  for transverse $\rho$ mesons. 
At the same time, a stronger modification of $A^L$ is exhibited. Thus, 
coupling iii) induces a qualitatively different
behaviour on $A_\rho$, removing much of the distinction between transverse and 
longitudinal $\rho$ mesons. The results at high momenta are shown in 
Fig. \ref{a_comp8t}. In this kinematical region spin-$\frac{5}{2}$
resonances, for which we have only a non-relativistic coupling at hand,  
have noticeable influence on the results. Therefore, by changing the 
interaction in the spin-$\frac{3}{2}$ sector and leaving the 
spin-$\frac{5}{2}$ sector unchanged, the effect of the coupling scheme
on the spectral function is underestimated. In order to obtain an upper limit
for this effect, we treat the spin-$\frac{5}{2}$ resonances the same
way as spin-$\frac{3}{2}$ resonances, correcting for the different spins 
of the states by a factor of $3/2$. This procedure is reasonable, since
--- except for the multiplicity --- the spin of the resonance affects 
mostly the angular dependence of the matrix element. We have checked that
for the transition from spin-$\frac{1}{2}$ to spin-$\frac{3}{2}$ resonances
our prescription works fine within $10 \%$.

The most distinct feature of coupling ii) is an increased resonance
contribution to $A^L$ at large momenta. The transverse channel, however, 
does not undergo large modifications as compared to coupling i).

One should keep in mind, that in general the actual coupling
of a resonance to the $\rho\,N$ channel might contain contributions from 
all three couplings. Naturally this will smear out the differences. 
One expects from the above discussion, that the general tendency will go 
towards a less modified transverse $\rho$ meson
than predicted from coupling i), whereas in the longitudinal
channel a stronger modification is likely.
It is beyond the scope of this work, however, to find realistic 
weighting factors for each coupling. This
can only be achieved by a complete relativistic partial wave analysis of 
$\pi\,N \rightarrow \pi\,\pi\,N$ data.


\section{Conclusions}
\label{conclu}

We have performed a calculation of the $\rho$ spectral function $A_\rho$ in nuclear
matter within a relativistic resonance model. The aim was, to put predictions
concerning the properties of $\rho$ mesons in nuclear matter, which were derived
in a non-relativistic reduction \cite{pp98}, on a more solid basis. Especially the 
momentum dependence of $A_\rho$ was examined. 
This discussion was motivated by the observation that the results 
of a non-relativistic calculation depend strongly on the reference frame in which
the non-relativistic vertex functions are evaluated.
We also investigated 
the influence of different resonance parameters. We argued, that
the resonance parameters as obtained in the work of Manley {\it et al} \cite{man1,man2} 
provide much more reliable information on $\rho\,N$ scattering  than 
the estimated resonance parameters listed in PDG \cite{pdg}.
Finally, we studied if and to which extent the obtained results  
depend on the structure of the chosen coupling. 
Therefore, various coupling schemes for the $RN\rho$ vertex
were employed in the spin-$\frac{3}{2}$ sector.

A very interesting finding of this work is that a non-relativistic approach
is in very good agreement with a relativistic one and should be good enough
for all practical purposes, provided that the vertex
functions for both the determination of the coupling constant and 
the scattering amplitude are evaluated in the cm-frame, i.e.
the rest frame of the intermediate resonance. In contrast, it is not
advisable to determine the coupling constant with cm-kinematics but the
scattering amplitude in the rest frame of the nucleon, as was done in
all previous works \cite{fp,pp98,rw,rw2}. We could demonstrate
that this procedure leads to dramatic deviations from a relativistic calculation.

As an important finding in our previous non-relativistic calculation, we 
reported on the dominant role of the $D_{13}(1520)$ in $\rho\,N$ scattering.
At low momenta these results are supported by the relativistic calculation. 
The parameters of this resonance are well known.
Furthermore, different vertex functions show only marginal effects on the results. 
For higher momenta the role of the $D_{13}(1520)$ is reduced in a 
relativistic calculation.

Another striking outcome of a non-relativistic calculation was the very different 
behaviour of $A_\rho^T$ and $A_\rho^L$ at high momenta. In the light of a
relativistic version of $A_\rho$ this distinction needs to be revised. 
As an overall effect, taking into account all modifications on the calculations,
a transversely polarized $\rho$ meson keeps
a larger amount of its total strength around the pole mass than indicated
in the earlier calculation. The previous finding
of a completely structureless strength distribution in the transverse channel 
at high momenta can therefore not be confirmed. In particular, the coupling with
a derivative acting on the $\rho$ field tensor produces a transverse spectral function
$A_\rho^T$, which undergoes less modifications at high momenta and is very
similar to $A_\rho^L$.

The reliability of the calculation in the transverse channel
is limited by poorly known resonance
parameters, especially their partial decay width into the $\rho\,N$ channel.
In particular, a  large uncertainty is connected with the branching 
ratio of the $P_{13}(1720)$.
In this partial wave, experimental data do not allow a conclusive determination
of the resonance parameters.

In the longitudinal channel only quantitative changes occur. Here, the results
are governed by physical constraints at threshold and at the photon point.
The change from a non-relativistic to a relativistic calculation
has no substantial impact on the results.
As in the transverse channel, at large momenta the coupling with a 
derivative acting on the $\rho$ field tensor, produces the largest 
quantitative change. In this case it leads to a pronounced broadening of the 
$\rho$ peak.
Up to which extent this coupling is present in the actual 
$RN\rho$ vertex remains an open question.
As a general tendency, it will lead to a smaller modification of the transverse 
channel, whereas the influence of resonance scattering for
longitudinal $\rho$ mesons is enhanced.

Altogether, we can confirm
the dominant influence of the $D_{13}(1520)$ on the propagation 
of $\rho$ mesons in nuclear matter at low momenta. 
The sensitivity to the chosen coupling scheme is small and the 
physical parameters of this resonance are well under control.
The distinction between $A_\rho^T $ and $A_\rho^L$
as found in our non-relativistic study 
also exists in the relativistic calculation, though to a reduced extent.

\section{Acknowledgments}

The authors would like to thank W. Peters for pointing out the problems
connected with the relativistic calculation of the spectral function. 
M.P. enjoyed many
discussion with M. Effenberger concerning the Manley analysis.


\begin{appendix}
\label{app}
\section{Appendix}

\subsection{$|{\cal M}_{RN\rho}^{T(L)}|^2$}
In this Appendix we show the problem arising from the completeness relation
Eq. \ref{compl_naive}.
The squared matrix element $|{\cal M}_{RN\rho}^{T(L)}|^2$ for the decay of a 
resonance with the quantum numbers $J^\pi=\frac12^+$
into $N\,\rho$ is readily calculated, using  
Eqs. \ref{matrix2}, \ref{lagrangian} and (erroneously) \ref{compl_naive}:
\beqn
\label{negma2}
|{\cal M}_{RN\rho}^T|^2 &\stackrel{?}{=}& 
-\frac{P^T_{\mu\nu}}{2}\,Tr\left[(\slash \hspace{-0.2cm}p_N+m_N)\,
{\cal V}^{\mu}_{\frac12}\,
(\slash \hspace{-0.2cm}k+m_R)\,{\cal V}^{\nu}_{\frac12} \right] 
\nonumber \\ \nonumber && \nonumber \\  
&=& 4\,I_\Gamma\,\left(\frac{f_{RN\rho}}{m_\rho}\right)^2\,m_N\,
\left(\omega^2\, (m_N + \omega - m_R) + \right. \nonumber
\\ && \hspace{3.6cm} \left.{\bf q}^2\,(m_N+m_R-\omega)    \right)  \nonumber \\
&& \\
|{\cal M}_{RN\rho}^L|^2 &\stackrel{?}{=}& 
-P^L_{\mu\nu}\,Tr\left[(\slash \hspace{-0.2cm}p_N+m_N)\,{\cal V}^{\mu}_{\frac12}\,
(\slash \hspace{-0.2cm}k+m_R)\,{\cal V}^{\nu}_{\frac12} \right]
\nonumber \\ \nonumber && \nonumber \\ 
&=& 4\,I_\Gamma\,\left(\frac{f_{RN\rho}}{m_\rho}\right)^2\,
m_N\,q^2\,\left(m_N + \omega -m_R \right)  \nonumber \qquad . 
\eeqn
The calculation is done in the rest frame of the nucleon, and
$I_\Gamma$ denotes an isospin factor, see Eq. \ref{iso}.
One easily verifies that  $|{\cal M}_{RN\rho}^{T/L}|^2$ can become negative.
At ${\bf q}=0$, for example, the squared matrix element changes sign
at $\omega = m_R - m_N$. Clearly, using Eq. \ref{complete}
instead of Eq. \ref{compl_naive} does not lead to such a pathological behaviour.

For resonances with negative parity we do not find such effects. 
This can be understood by observing that the problem arises 
at the on-shell point $\sqrt s = m_R$.
In the case of $s$-wave resonances a sign change at this point is not possible
however, since the amplitude is non-vanishing at the on-shell point.
This, of course, does not imply that treating the $s$-wave resonances with 
the unmodified propagator is correct.

\subsection{The Propagator}

According to \cite{bjdr} the propagator
$D_{\frac{1}{2}}(k)$ has to satisfy a number of 
constraints. Expressing the propagator in 
a Lehmann representation:  

\beq
\label{lehmann}
D_{\frac{1}{2}}(k) = \int_0^\infty\, dM^2\, \left[ \slash \hspace{-0.2cm}k \, 
\rho_{sp}(M^2) + M\,\rho_{sc}(M^2) \right] \, \frac{1}{k^2-M^2+i\epsilon}  \qquad ,
\eeq
these constraints translate into properties of the 
spinor and scalar spectral weight functions of the propagator,
$\rho_{sp}$ and $\rho_{sc}$ respectively:

\beq
\begin{array}{rl}
\label{condition}
\mbox{i)} & \mbox{$\rho_{sp}(M^2)$ and $\rho_{sc}(M^2)$ are both real} \\ \\
\mbox{ii)} & \rho_{sp}(M^2) > 0 \\ \\
\mbox{iii)} & \rho_{sp}(M^2)-\rho_{sc}(M^2) \ge 0  \qquad .\\  \\
\end{array}
\eeq

It is now possible to check whether the often used
expression from Eq. \ref{naive} is compatible
with these conditions. 
%
The comparison of Eqs. \ref{naive} and \ref{lehmann} allows the identification
of the functions $\rho_{sp}$ and $\rho_{sc}$:
\beqn
\label{rhos}
\rho_{sp}^{trial}(k^2) &=& \frac{1}{\pi}\frac{\sqrt{k^2}\,\Gamma}
{(k^2-m_R^2)^2 + k^2\,\Gamma^2} \nonumber \\
            &&                                                                    
\nonumber \\
\rho_{sc}^{trial}(k^2) &=& \frac{1}{\pi}\frac{m_R\,\Gamma}{(k^2-m_R^2)^2 + k^2\,\Gamma^2}  
                \\
            &&                                                                   
 \nonumber \\
            &=& \rho_{sp}^{trial}(k^2)\,\frac{m_R}{\sqrt{k^2}}
            \qquad . \nonumber 
\eeqn
Clearly $\rho_{sp}^{trial}$ and $\rho_{sc}^{trial}$, 
as obtained from Eq. \ref{naive},
violate the last condition in Eq. \ref{condition} for $k^2<m_R^2$.
 
In order to construct a more reasonable propagator which incorporates the
constraints from Eq. \ref{condition}, we will choose the spectral 
weight functions such that they fulfill these constraints and then 
calculate the full propagator directly from Eq. \ref{lehmann}.
Keeping things as simple as possible we make the following {\it ansatz} for 
$\rho_{sp}$ and $\rho_{sc}$:
\beq
\rho_{sp} = \rho_{sc} = \rho_{sp}^{trial} \qquad .
\eeq 

This is readily done for the imaginary part of 
the propagator, leading to the expression:
\beq
\label{imbetter}
\mbox{Im }D_\frac{1}{2}(k) = \,\mbox{Im }\left(\frac{\slash\hspace{-.2cm}k + 
\sqrt{k^2}}{k^2-m_R^2+ i\,\sqrt{k^2}\,\Gamma} \right) \quad .
\eeq

For the real part of $D_\frac{1}{2}(k)$, the integral can be solved analytically only, if 
a constant decay width $\Gamma$ is assumed, leading to a lengthy expression.
Taking into account the mass dependence of the decay width
makes a numerical computation mandatory.
Concerning the calculation of the $\rho$ meson self energy
we checked by explicit comparison that the {\it ansatz}
\beq
\label{rebetter}
\mbox{Re } D_\frac{1}{2}(k) = \,\mbox{Re }\left(\frac{\slash\hspace{-.2cm}k + 
\sqrt{k^2}}{k^2-m_R^2+i\,\sqrt{k^2}\,\Gamma} \right) 
\eeq
produces a satisfactory agreement with the exact numerical solution.
Small deviations can be seen, but these have virtually no effect
on the spectral function of the $\rho$ meson. Obviously, Eqs. \ref{imbetter}
and \ref{rebetter} lead back to our guess Eq. \ref{better}.

One can write down a Lehmann representation for the propagator of spin-$\frac32$ particles
$D_\frac{3}{2}^{\alpha\beta}(p)$ in a
straightforward extension of Eq. \ref{lehmann}. This expression contains
in principle eight spectral weight functions, four multiplying the spinor part
of $D_\frac{3}{2}^{\alpha\beta}(p)$ and four its scalar part. 
By analogy to
the case of spin-$\frac{1}{2}$ fields, we require them to be identical and put them equal
to $\rho_{sp}^{trial}$ from Eq. \ref{rhos}. Thus, we obtain
the result from Eq. \ref{32better}.


\subsection{Results for $\Omega^{T/L}$}
In this Appendix we give explicit expressions for the quantities $\Omega^{T/L}$ which
have been introduced in Sect. \ref{results}. They are evaluated in the rest frame of 
nuclear matter, thus $\omega$ and ${\bf q}$ refer to energy and momentum of the 
$\rho$ meson in the lab-frame and $q^2$ denotes its squared invariant mass. The invariant
energy of the intermediate resonance $\sqrt s$ is given by $\sqrt s = q^2+m_N^2+2m_N\omega$.
Note that in the spin-$\frac32$ sector singular terms $\sim \frac{1}{s}$ show up, 
in accordance with the discussion in Appendix 2. Note that all results need to 
be multiplied by a factor $\frac{f_{RN\rho}^2}{m_\rho^2}$.
We first list the results obtained from the relativistic Lagrangians in 
Eq. \ref{lagrangian}:

\beq
\label{r1}
\begin{array}{|lll|}\hline && \\
J^\pi=\frac12^+&:\hspace{2.5cm}&
\Omega^L = 4\,m_N
\,q^2\,\left(m_N+\omega-\sqrt s \right) \\ &&\\
&&\Omega^T = \Omega^L + 8\,m_N^2\,{\bf q}^2 \\ \hline&& \\ 

J^\pi=\frac12^-&:&
\Omega^L = 4\,m_N\,q^2\,\left(m_N+\omega+\sqrt s \right) \\&&\\&&
\Omega^T = \Omega^L + 8\,m_N^2\,{\bf q}^2 \\ \hline&& \\

J^\pi=\frac32^+&:&
\Omega^L = \frac83\,m_N\,q^2\,\left(m_N+\omega-\sqrt s \right) \\&&\\&&
\Omega^T = \Omega^L + \frac83\,m_N^2\,{\bf q}^2\,\left(1+\frac{m_N^2+m_N\,\omega}{s}\right)
\\ \hline&& \\

J^\pi=\frac32^-&:&
\Omega^L = \frac83\,m_N\,q^2\,\left(m_N+\omega+\sqrt s \right) \\&&\\&&
\Omega^T = \Omega^L + \frac83\,m_N^2\,{\bf q}^2\,\left(1+\frac{m_N^2+m_N\,\omega}{s}\right)
\\\hline
\end{array}
\eeq
Using the coupling which is denoted by ii) in Sect. \ref{uncertain} one derives: 
\beq
\label{r2}
\begin{array}{|lll|} \hline &&\\
J^\pi=\frac32^+&:\hspace{1cm}&
\Omega^L = \frac83\,m_N^3\,q^2\,\left(m_N+\omega-\sqrt s \right)\,
\left(1-\frac{{\bf q^2}}{s}\right)
\\&&\\&&
\Omega^T = \frac83\,m_N^3\,\omega^2\,\left(m_N+\omega-\sqrt s \right)\\\hline&&\\ 
J^\pi=\frac32^-&:\hspace{1cm}&
\Omega^L = \frac83\,m_N^3\,q^2\,\left(m_N+\omega+\sqrt s \right)\,
\left(1-\frac{{\bf q^2}}{s}\right)
\\&&\\&&
\Omega^T = \frac83\,m_N^3\,\omega^2\,\left(m_N+\omega+\sqrt s \right)\\\hline 
\end{array}
\eeq
The coupling denoted by iii) in Sect. \ref{uncertain} leads to:
\beq
\label{r3}
\begin{array}{|lll|} \hline &&\\
J^\pi=\frac32^+&:\hspace{1cm}&
\Omega^L = \frac83\,m_N\,q^4\,\left(m_N+\omega-\sqrt s \right)\,
\left(1-\left(\frac{m_N^2}{q^2}\right) \frac{{\bf q^2}}{s}\right)\\&&\\&&
\Omega^T = \frac83\,m_N\,q^4\,\left(m_N+\omega-\sqrt s \right)\\\hline&&\\ 

J^\pi=\frac32^-&:\hspace{1cm}&
\Omega^L = \frac83\,m_N\,q^4\,\left(m_N+\omega+\sqrt s \right)\,
\left(1-\left(\frac{m_N^2}{q^2}\right) \frac{{\bf q^2}}{s}\right)\\&&\\&&
\Omega^T = \frac83\,m_N\,q^4\,\left(m_N+\omega+\sqrt s \right)\\\hline 

\end{array}
\eeq

For comparison we also give the non-relativistic results \cite{pp98}. 
Note the appearance 
of an additional factor $\sqrt s + m_R$, which needs to be introduced since
in Eq. \ref{spur} the scattering amplitude is formulated using a relativistic propagator.

\beq
\label{nr}
\begin{array}{|lll|} \hline &&\\
J^\pi=\frac12^+&:\hspace{2cm}&
\Omega^L = 0.\\&&\\&&
\Omega^T = 4\,m_N\,{\bf q}^2\,\left(m_R+\sqrt s\right)\\\hline&&\\ 

J^\pi=\frac12^-&:&
\Omega^L = 4\,m_N\,q^2\,\left(m_R+\sqrt s \right) \\&&\\&&
\Omega^T = \Omega^L + 4\,m_N\,{\bf q}^2\,\left(m_R+\sqrt s\right)\\ \hline &&\\

J^\pi=\frac32^+&:&
\Omega^L = 0.\\&&\\&&
\Omega^T = \frac83 \,m_N\,{\bf q}^2\,\left(m_R+\sqrt s\right)
\\  \hline &&\\

J^\pi=\frac32^-&:&
\Omega^L = \frac83\,m_N\,q^2\,\left(m_R+\sqrt s \right) \\&&\\&&
\Omega^T = \Omega^L + \frac83 \,m_N\,{\bf q}^2\,\left(m_R+\sqrt s\right)
\\ \hline && \\ 

J^\pi=\frac52^+&:&
\Omega^L = 0. \\&&\\&&
\Omega^T = \frac45 \,m_N\,{\bf q}^2\,\left(m_R+\sqrt s\right) \\\hline

\end{array}
\eeq
Here ${\bf q}$ refers to either the cm-momentum or the lab-momentum of the $\rho$ meson,
depending on which kinematic has been chosen.

\end{appendix}


\newpage


\begin{table}
\begin{tabular}{|l||c|c|c|c|l|l|} \hline
       & $m_R$(GeV) & $\Gamma_{tot}$(MeV) & $\Gamma_{N\rho}$(MeV) & 
       $\Gamma_{N\rho}^{PDG}$(MeV) & $f_{RN\rho}$ & \\ \hline \hline
$P_{33}(1232)$ & 1.231 & 118 & --   & --   & 13.4 & $p$-wave \\ \hline
$D_{13}(1520)$ & 1.524 & 124 & 26 & 24 & 6.5 & $s$-wave \\ \hline
$S_{31}(1620)$ & 1.672 & 154 & 44 & 20 & 2.0 &$s$-wave    \\ \hline
$S_{11}(1650)$ & 1.659 & 173 & 5  & 15 & 0.5  & $s$-wave    \\ \hline
$F_{15}(1680)$ & 1.684 & 139 & 11 & 12 & 6.46  & $p$-wave   \\  \hline
$D_{33}(1700)$ & 1.762 & 599 & 46 & 120 & 1.9  & $s$-wave   \\ \hline
$P_{13}(1720)$ & 1.717 & 383 & 333 & 100 & 10.1  &$p$-wave     \\ \hline
$F_{35}(1905)$ & 1.818 & 327 & 282 & 210 & 17.5  &$p$-wave  \\ \hline \hline
$P_{13}      $ & 1.879 & 498 & 217 & --    & 3.1  & $p$-wave   \\ \hline
$D_{33}(1940)$ & 2.057 & 460 & 162 & --    & 0.5  & $s$-wave   \\ \hline
$F_{15}(2000)$ & 1.903 & 494 & 296 & --    & 9.5  & $p$-wave   \\ \hline
$D_{13}(2080)$ & 1.804 & 447 & 114 & --    & 1.4  & $s$-wave   \\ \hline
$S_{11}(2090)$ & 1.928 & 414 & 203 & --    & 0.7  & $s$-wave   \\ \hline

\end{tabular}
\caption{
\label{barres} List of the resonances included in this calculation. The parameters
in the first three columns are taken from the Manley analysis \cite{man1,man2}. In the
fourth column the resonance decay width into $\rho\,N$ as presented in 
the PDG \cite{pdg} is 
listed. The fifth column shows the coupling strength of the resonances to $\rho\,N$
as obtained from expression (\ref{coupldet}). In the last column we denote
if the respective resonance decays in a relative $s$-wave or $p$-wave into
the $\rho\,N$ channel.
The resonances
below the double line have not been included in our previous analysis \cite{pp98}.}
\end{table}

\newpage


\begin{figure} 
\begin{center}
\epsfig{file=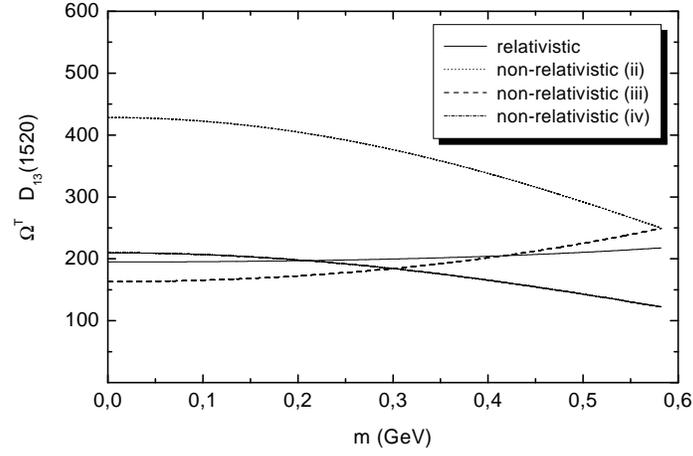,width=10cm}
\epsfig{file=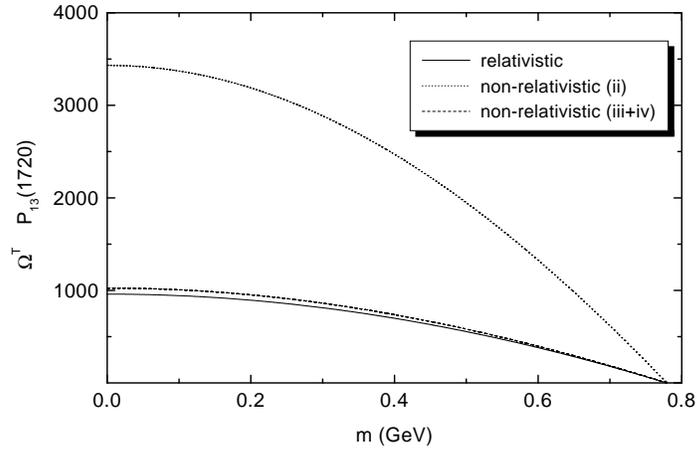,width=10cm}
 \caption{ 
\label{spurt} $\Omega^T$ for the $D_{13}(1520)$ resonance (upper graph)
and the $P_{13}(1720)$ (lower graph) at $s=m_R^2$ as a function of the invariant
mass $m$ of the $\rho$ meson.
Shown are three different calculations, which
are explained in Sect. \ref{results}.}
\end{center}
\end{figure}


\end{document}